\documentclass[english]{article}
\usepackage[T1]{fontenc}
\usepackage[latin9]{inputenc}
\pdfoutput=1
\usepackage{geometry}
\usepackage{color}
\usepackage{float}
\usepackage{calc}
\usepackage{textcomp}
\usepackage{mathrsfs}
\usepackage{amsmath}
\usepackage{amsthm}
\usepackage{amssymb}
\usepackage{graphicx}
\usepackage[bottom]{footmisc}
\usepackage{multirow}

\makeatletter

\providecommand{\tabularnewline}{\\}

\usepackage{thmtools}

\declaretheorem[style=definition,name=Theorem,qed=$\blacksquare$]{theorem}
\theoremstyle{definition}
\newtheorem{thm}{\protect\theoremname}
\theoremstyle{remark}
\newtheorem{rem}[thm]{\protect\remarkname}

\usepackage{ dsfont }
\usepackage[all]{xy}
\usepackage{cite}
\usepackage{colonequals}

\usepackage{textgreek}

\usepackage{tikz}

\makeatletter
\let\@fnsymbol\@arabic
\makeatother
\usetikzlibrary{patterns,snakes}
\usetikzlibrary{patterns,decorations.pathreplacing}
\usetikzlibrary{arrows.meta}
\usepackage{pgfplots}
\usetikzlibrary{arrows,decorations.markings,decorations.text}

\RequirePackage{mathtools}
\global\long\def\defi{\vcentcolon =}

\AtBeginDocument{

}

\makeatother

\usepackage{babel}
\providecommand{\remarkname}{Remark}
\providecommand{\theoremname}{Theorem}

\begin{document}
\global\long\def\u{u}
\global\long\def\p{P}
\global\long\def\X{X}
\global\long\def\me{\mu_{e}}
\global\long\def\sym{\textrm{sym}}
\global\long\def\grad{\nabla}
\global\long\def\le{\lambda_{e}}
\global\long\def\tr{\textrm{tr}}
\global\long\def\mc{\mu_{c}}
\global\long\def\skew{\textrm{skew}}
\global\long\def\curl{\textrm{Curl}}
\global\long\def\ac{\alpha_{c}}
\global\long\def\B{\mathscr{B}}
\global\long\def\R{\mathbb{R}}
\global\long\def\fr{\rightarrow}
\global\long\def\Q{\mathcal{Q}}
\global\long\def\A{\mathscr{A}}
\global\long\def\L{\mathscr{L}}
\global\long\def\D{\mathscr{D}}
\global\long\def\id{\mathds{1}}
\global\long\def\ds{\textrm{dev sym}}
\global\long\def\sph{\textrm{sph}}
\global\long\def\eg{\boldsymbol{\varepsilon}}
\global\long\def\aa{\boldsymbol{\alpha}}
\global\long\def\o{\boldsymbol{\omega}}
\global\long\def\ege{\boldsymbol{\varepsilon}_{e}}
\global\long\def\egp{\boldsymbol{\varepsilon}_{p}}
\global\long\def\punto{\,.\,}
\global\long\def\so{\mathfrak{so}\left(3\right)}
\global\long\def\Sym{\textrm{Sym}\left(3\right)}
\global\long\def\MM{\boldsymbol{\mathfrak{M}}}
\global\long\def\C{\mathbb{C}}
\global\long\def\gl{\mathfrak{gl}\left(3\right)}
\global\long\def\P{P}
\global\long\def\Lin{\textrm{Lin}}
\global\long\def\D{\boldsymbol{D}}
\global\long\def\a{\alpha}
\global\long\def\b{\beta}
\global\long\def\lle{\lambda_{e}}
\global\long\def\ce{\mathbb{C}_{e}}
\global\long\def\cm{\mathbb{C}_{\textrm{micro}}}
\global\long\def\cc{\mathbb{C}_{c}}
\global\long\def\axl{\textrm{axl}}
\global\long\def\dev{\textrm{dev}}
\global\long\def\Ls{\widehat{\mathbb{L}}}
\global\long\def\mh{\mu_{\textrm{micro}}}
\global\long\def\lh{\lambda_{\textrm{micro}}}
\global\long\def\vau{\omega_{1}^{int}}
\global\long\def\vad{\omega_{2}^{int}}
\global\long\def\axl{\textrm{axl}}
\global\long\def\Le{\mathbb{L}_{e}}
\global\long\def\Lc{\mathbb{L}_{c}}
\global\long\def\V{\mathbb{V}}
\global\long\def\as{\dagger}
\global\long\def\dd{\:\overset{{\scriptscriptstyle \triangledown}}{{\scriptscriptstyle \vartriangle}}\,}
\global\long\def\cM{\mathbb{C}_{\textrm{macro}}}
\global\long\def\mum{\mu_{\textrm{macro}}}
\global\long\def\lam{\lambda_{\textrm{macro}}}

\title{\vspace{-3cm}\textbf{Identification of scale-independent material parameters in the relaxed micromorphic model through model-adapted first order homogenization}}

\author{
Patrizio Neff\,\thanks{Patrizio Neff, corresponding author, patrizio.neff@uni-due.de, Head of Chair for Nonlinear
	Analysis and Modelling, Fakultät für Mathematik, Universität Duisburg-Essen,
	Mathematik-Carrée, Thea-Leymann-Straße 9, 45127 Essen, Germany},$\;$$\;$\, Bernhard Eidel\thanks{Bernhard Eidel, bernhard.eidel@uni-siegen.de, Universität Siegen,
	Institut für Mechanik, Heisenberg-group, Paul-Bonatz-Straße 9-11, 57076
	Siegen, Germany} ,\, Marco Valerio d\textquoteright Agostino\thanks{Marco Valerio d'Agostino,  marco-valerio.dagostino@insa-lyon.fr,
GEOMAS, INSA-Lyon,  20 avenue Albert Einstein,
69621, Villeurbanne, France},
\, and Angela Madeo\thanks{Angela Madeo, angela.madeo@insa-lyon.fr, GEOMAS, INSA-Lyon,  20 avenue Albert Einstein, 69621, Villeurbanne, France}
\textit{}\linebreak{}
\textit{}\\
\textit{}\\
\textit{Dedicated to Prof. Dr. Dr. h. c. Hans-Dieter Alber on the occasion }
	\\
	\textit{ of his 70. birthday with great esteem.}}
\maketitle\vspace{-7mm}
\begin{abstract}
We rigorously determine the scale-independent short range elastic parameters in the relaxed micromorphic generalized continuum model for a given periodic microstructure. This is done using both classical periodic homogenization and a new procedure involving the concept of apparent material stiffness of a unit-cell under affine Dirichlet boundary conditions and Neumann's principle on the overall representation of anisotropy. We explain our idea of "maximal" stiffness of the unit-cell and use state of the art first order numerical homogenization methods to obtain the needed parameters for a given tetragonal unit-cell. These results are used in the accompanying paper \cite{d2019effective} to describe the wave propagation including band-gaps in the same tetragonal metamaterial.
\end{abstract}
\addtocounter{footnote}{5} \vspace{3mm}

\vspace{-1mm}
\textbf{}
(effective constitutive equations).

\vspace{-3mm}
\tableofcontents{}

\section{Introduction}

\addtocounter{footnote}{-5} 

In this work we aim to partially homogenize a periodic linear elastic Cauchy material towards a relaxed micromorphic continuum. Classical homogenization is a mature subject and delivers rigorously the effective linear elastic stiffness tensor $\cM$, which then describes the (very) large scale response of the periodic structure in arbitrary boundary value problems. The classical homogenization approaches are valid when the characteristic size of the studied microstructure is orders of magnitudes smaller than the characteristic size of the structure  (scale-separation hypothesis). The effective stiffness tensor $\cM$ can be conveniently characterized by looking at any possible unit-cell of the periodic Cauchy material and applying periodic boundary conditions (PBC)\footnote{As is well-known, under affine loading, the response of a large periodic structure is periodic up to a vanishing boundary layer.}. The ensuing FEM-problem is nowadays routinely solved \cite{schroder2014numerical}.
\begin{figure}[H]
	\begin{centering}
		\includegraphics[scale=0.35]{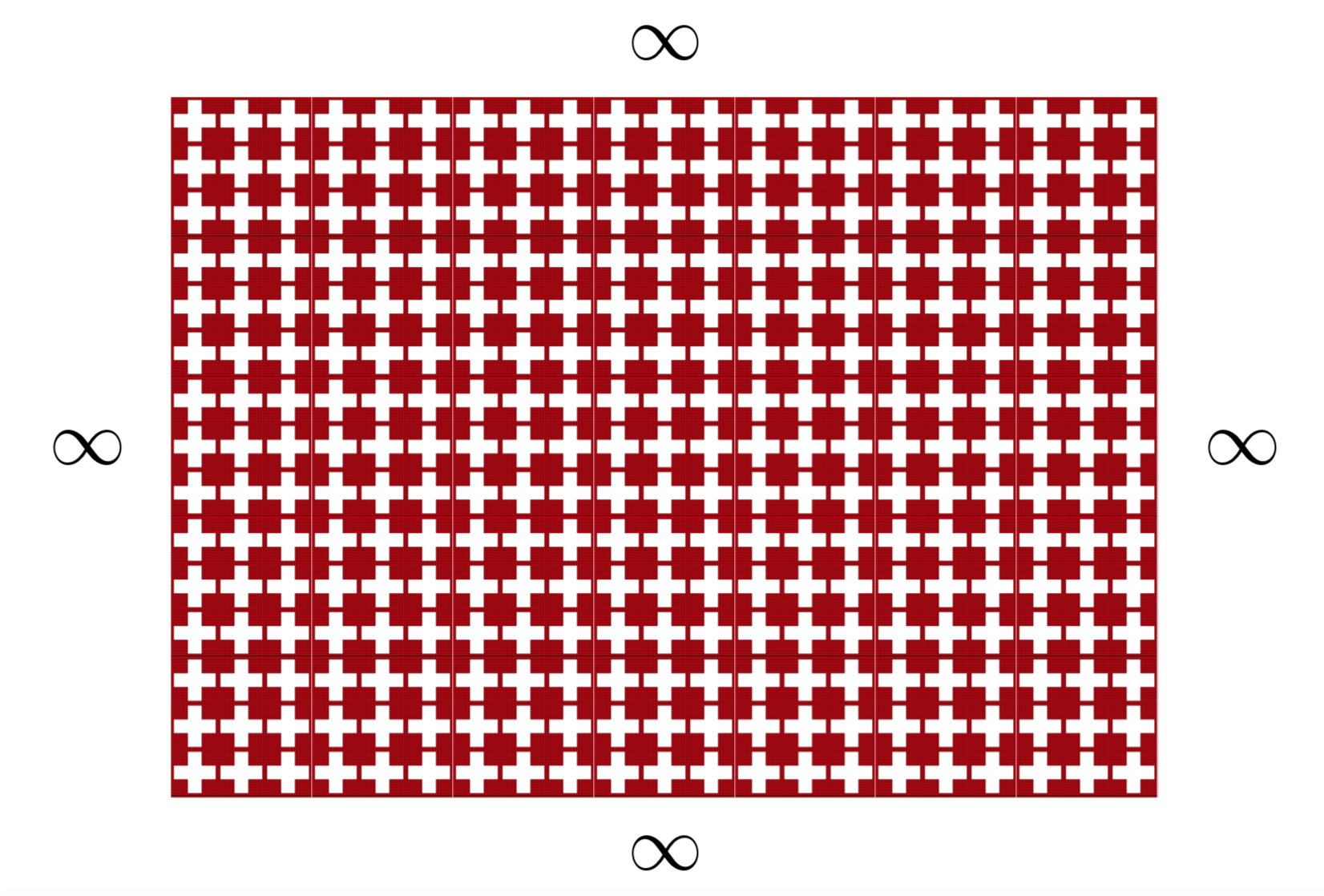}\caption{Infinite periodic structure. The effective properties are given by the effective stiffness tensor $\cM$ which can be obtained rigorously by $\Gamma$-convergence and asymptotic homogenization.}
		\par\end{centering}
\end{figure}
However, giving up the scale-separation hypothesis, if one is interested in calculating a finite sized sample of the periodic structure under general non-affine loadings, it becomes apparent that its response is hardly governed by a homogeneous linear elastic surrogate model \cite{sridhar2016homogenization,pham2013transient}. Here, higher order models \cite{smyshlyaev2000rigorous} or extended continuum models come into play. A key feature of these models is that they are able to naturally describe the appearing size-effects (typically "smaller is stiffer").
\begin{figure}[H]
	\begin{centering}
		\includegraphics[scale=0.45]{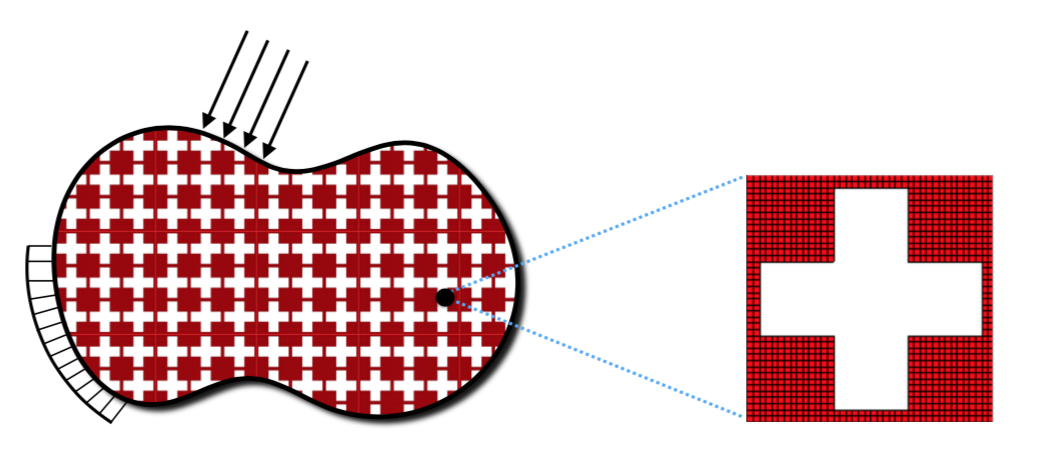}\caption{Finite sized sample of a periodic structure. The scale-separation hypothesis does not anymore apply.}
		\par\end{centering}
\end{figure}
This ability is either due to the incorporation of higher gradients of displacement or by amending the kinematics on the homogenized scale with suitable independent additional fields for which new balance equations have to be devised. This leads us to consider the micromorphic framework, introduced by Eringen and Mindlin \cite{eringen1964nonlinear,eringen1966mechanics,eringen1999microcontinuum,mindlin1964micro} following earlier work of the Cosserat brothers \cite{cosserat1909theorie}. At present, a rigorous asymptotic expansion or a $\Gamma$-convergence result \cite[Theorem 5.6]{braides2006handbook} towards a homogenized model for this transition scale seems to be out of reach for periodic materials showing a high stiffness contrast of their phases as in the present case and other technical difficulties. However, heuristically and computationally \cite{rokovs2019micromorphic,pham2013transient,sridhar2016homogenization,kouznetsova2002multi,kouznetsova2004multi}, the micromorphic framework seems to be well suited in these circumstances. In the absence of any comprehensive rigorous result (but see \cite{smyshlyaev2000rigorous}), we therefore postulate for this work that the suitably homogenized model will be of the micromorphic type.

A serious drawback of these models  is the increasing number of unknown material parameters, that have to be identified as well as the problematic interpretation of the new kinematical fields. 

It is therefore mandatory to devise homogenization rules (analytical or computational) in order to determine relevant material parameters. While the question of homogenization towards an extended continuum model has seen a lot of effort in the last two decades (championed by Samuel Forest and his group \cite{forest2002homogenization,forest1998mechanics,forest2011generalized}, recently also Hütter \cite{hutter2017homogenization,hutter2019micro}) it is fair to say that no universally valid answer has been found.

\bigskip

\noindent The performed heuristic homogenization procedures typically involve the following steps:
\begin{description}
	\item[a)] a postulate on the form of the extended balance equations for generalized macroscopic stresses,
	\item[b)] a postulate on the relation between microscopic Cauchy stresses and generalized macroscopic stresses, without specifying the precise macroscopic constitutive relations (e.g., a priori restrictions on material symmetry on the micro or macro-scale are not taken into account),
	\item[c)] a generalized Hill-Mandel postulate expressing energy equivalence between the micro and the macro-scale, 
	\item[d)] a postulate on the connection between microscopic displacements and the additional macroscopic fields\footnote{For the micromorphic model, postulate d) implies a direct interpretation of what the new degrees of freedom (the non symmetric micro-distortion $P\in\R^{3\times3}$) is. While we do not discard such a direct micro-macro relation, we rather believe that any simple relation will fall short of the truth for the relaxed micromorphic model.},
	\item[e)] use of higher order boundary conditions on the unit-cell level for triggering inhomogeneous response  and activating the higher order effects. 
\end{description}
The combination of these 5 steps can involve analytical and computational procedures and leads to some form of "homogenized" material properties. 
Step a) seems to be uncritical, since most authors working in the field of extended continuum mechanics agree on the same or similar format of balance relations \cite{ehlers2018particle,hutter2017homogenization,hutter2019micro,mindlin1964micro,eringen1966mechanics,eringen1999microcontinuum}. Step b) already implies some averaging rules \cite{hutter2017homogenization} and is therefore open to discussion. Step c) is a  generalization of the procedure in homogenization towards linear elasticity but involves some arbitrariness, since the proper form of boundary conditions has not yet been found. Step d) seems to be essential but no general agreement has been reached. For example \cite{forest1999aufbau,forest2002homogenization,forest2011generalized} identifies the micro-distortion $P\in\R^{3\times3}$ (the additional kinematical field) with the first moment of the microscopic displacement field on the unit-cell level.  Typically the micro-distortion cannot be prescribed by a boundary condition at the micro-scale in contrast to the classical approach.  Hütter \cite{hutter2017homogenization,hutter2019micro} follows Forest \cite{forest2002homogenization,forest1998mechanics,forest2011generalized,trinh2012evaluation} but introduces in addition a weighting function. In classical first order homogenization, d) is a consequence of c). Finally, step e) seems to be a natural consequence of the procedure to close the argument and has been adopted by Gologanu et al. \cite{gologanu1997recent}, Kouznetsova et al. \cite{kouznetsova2002multi,kouznetsova2004multi},  Forest and Sab \cite{forest1998cosserat}, Forest and Trinh \cite{forest2011generalized}, Jänicke \cite{janicke2009two} and Diebels et al. \cite{diebels2003stress}.

\bigskip

Let us point out the problematic features of the sketched procedure. There are a number of natural requirements the homogenization process should ideally satisfy:

\begin{description}
	\item[1.)] if the material on the micro-scale is homogeneous and linear elastic, the homogenization should leave invariant the response, i.e., homogeneous Cauchy elasticity on the micro-scale turns into the same Cauchy response on the macro-scale,
	
	\item[2.)] there should be a clear separation between (short-range) scale-independent parameters (e.g., a classical shear modulus) and (long-range) scale-dependent parameters (e.g., a characteristic length scale) in the homogenized model,
	\item[3.)] for very large sample sizes the response should still be governed by linear elasticity with effective stiffness tensor $\cM$, and this should induce useful relations between the scale-independent material parameters of the micromorphic model. Here, we call \textbf{scale-independent material parameters} those coefficients in the extended continuum model that uniquely determine the large scale response,
	\item[4.)] the stiffness of the homogenized model should be bounded irrespective of the sample size (as the input micro-scale stiffness is certainly bounded). In other words, the possible storage of elastic energy in arbitrarily small windows on the homogenized scale should be bounded for all non-affine boundary conditions. This avoids unstable parameter identification in a series of size-experiments \cite{neff2010stable},
	\item[5.)] the homogenized parameters need to be true material parameters, independent of the applied loading and not influenced by boundary layer effects \cite{madeo2016new}.
\end{description}

At present, it is not known whether requirement 1.) can be satisfied by available homogenization rules, see \cite{hutter2019micro} (for example the  approach of Forest seems not to be suitable). Requirement 2.) is usually not addressed but a closer look reveals that it may be violated in the Cosserat framework. Requirement 3.) seems to be mostly ignored and we will indicate that it cannot be met in the general micromorphic setting. Requirement 4.) has led to restrictions in the Cosserat model for bending and torsion \cite{neff2010stable}. In general, it cannot be satisfied for second gradient continua. Finally, requirement 5.) is difficult to establish. In \cite{madeo2016new,ghiba2016variant} it is shown that parameters in second gradient  formulations may be boundary value dependent due to the presence of nontrivial null-Lagrangians. Thus, 4.) and 5.) seem to rule out second gradient formulations for our purpose.

\vspace{5mm}

In this work we want to deliberately dispense with the postulates b), c) and d) and instead place ourselves in a variational context and postulate the extended micromorphic kinematics together with a postulate on the suitable form of the free energy in an effort to reduce the complexity and intrinsic problems of the general Eringen-Mindlin micromorphic model \cite{romano2016micromorphic}. We have called our model "relaxed" micromorphic model. Our goal is to perform an identification of the (short-range) scale-independent material parameters, consistent with the requirements 1)$\ldots$5) towards homogenization.
The rational of the relaxed micromorphic model will guide us and provide some surprising novel routes.
Notably, the scale-independent material parameters will be determined solely with standard methods of first order homogenization which is sensible for the described short-range interaction.

\subsection{The static relaxed micromorphic model}

The relaxed micromorphic model is a generalized continuum model which includes a characteristic size. For the static case, it can be written in a  variational framework. The goal is to find the macroscopic mean displacement $u:\Omega\subseteq\R^3\fr\R^3$ and the non-symmetric micro-distortion field $P:\Omega\subseteq\R^3\fr\R^{3\times3}$ minimizing
\begin{equation}\label{ener1}
\int_\Omega W\left(\nabla u,P,\textrm{Curl} P \right) -\left\langle f,u \right\rangle   dx\;\longrightarrow\;\min.\qquad(u,P)\in H^1(\Omega)\times H(\curl)
\end{equation}
where the energy is represented as (see \cite{neff2014unifying,barbagallo2016transparent,madeo2014band,madeo2017modeling,madeo2016reflection})
\begin{align}
W\left(\nabla\u,\P,\curl\,\p\right) & =\underbrace{\frac{1}{2}\left\langle \mathbb{C}_{e}\,\sym\left(\grad\u-P\right),\sym\left(\grad\u-P\right)\right\rangle _{\R^{3\times3}}}_{\textrm{anisotropic elastic - energy}}+\underbrace{\frac{1}{2}\left\langle \mathbb{C}_{\textrm{micro}}\,\sym\,P,\sym\,P\right\rangle _{\R^{3\times3}}}_{\textrm{micro - self - energy}}\nonumber \\
& \qquad+\underbrace{\frac{1}{2}\left\langle \cc\,\skew\left(\grad\u-P\right),\skew\left(\grad\u-P\right)\right\rangle _{\R^{3\times3}}}_{ \substack{\text{invariant local fourth-order anisotropic} \\ \text{rotational elastic coupling}}}+\underbrace{\frac{\mu\,L_{c}^{2}}{2}\left\|\curl\,P\right\|^2}_{\textrm{relaxed curvature}}.\label{eq:bd33}
\end{align}

Here, $\ce,\,\cm$ are standard positive definite elasticity tensors with minor and major symmetries mapping symmetric matrices to symmetric matrices, $\cc$ is a positive semi-definite rotational coupling tensor mapping skew-symmetric matrices to skew-symmetric matrices, $L_c\geq0$ is a characteristic length scale and $\mu$ is a typical effective shear modulus\footnote{For the presentation we have chosen throughout the simplest representation of the curvature energy - a one constant isotropic format. }. The $\curl-$operator $\curl\,P$ acts row-wise on $P$. In all discussed models, linearized frame-indifference dictates that the scale-independent contribution of $P$ can only occur through a dependence on $\sym\,P$.  This framework is based on the additive split 
\begin{equation}\label{curl coupling}
\nabla u= \textcolor{red}{e}+\textcolor{blue}{P},\qquad 0=\curl\, \textcolor{red}{e}+\curl\,\textcolor{blue}{P}
\end{equation}
of the total displacement gradient into meso-scale and micro-scale contributions, $e$ and $P$ respectively. While $\nabla u $ is compatible by definition, $e$ and $P$ are in general incompatible, i.e., not a gradient of a vector-field.  Note that $\nabla u$ and $P$
are still macroscopic variables\footnote{Despite the name micromorphic model.}. The role of $P$ is twofold. On the one hand it describes the collective interaction in a cluster of unit-cells, e.g., $2\times2$ or $3\times3$ unit-cells, i.e., the interaction between unit-cells up to a certain range. On the other hand, $P$ also influences the short-range stiffness. Thus $P$ mainly represents a long-range fluctuation field giving rise to higher order moment stresses in contrast to $e=\nabla u-P$, which describes the remaining short-range elastic interaction in and between neighboring unit-cells (the unit-cells represent the macroscopic points of the homogenized continuum only under the scale-separation hypothesis) leading to the force-stresses. Missing is the micro-fluctuation (inside a unit-cell). This micro-fluctuation is conceptionally averaged out in the micromorphic model. In this view, (depending on the characteristic length $L_c$) we claim that the micro-distortion $P$ is not necessarily related to some average of micro-displacements in the unit-cell (as done by Forest \cite{forest2002homogenization,forest1998mechanics,forest2011generalized}, Hütter \cite{hutter2017homogenization,hutter2019micro}, Biswas and Poh\cite{biswas2017micromorphic} among others) but $P$ is a fully nonlocal object and truly an independent field not slave to the displacement\footnote{The situation is different when one considers homogenization towards a second gradient continuum (or micromorphic approximations thereof) where there is no independent kinematical field. More precisely, the case $\widehat{\C}_e\gg1$, $L_c\ll1$ would be consistent with determining $P$ as some average of the micro-displacements over a unit-cell.}.
\begin{figure}[H]\label{perturbation}
    \begin{centering}
    	\includegraphics[scale=0.4]{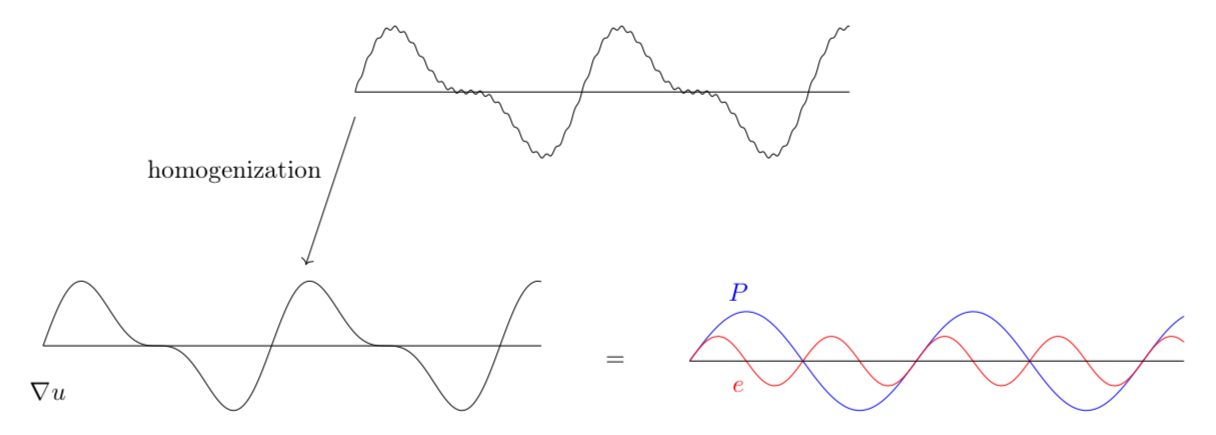} 		
    	\caption{Long-range fluctuation field $P$, short-range elastic interaction scale $e$ and averaged out micro-fluctuation.}
    	\par\end{centering}
\end{figure}
The static equilibrium equations are the Euler-Lagrange equations to \eqref{ener1} and \eqref{eq:bd33}, which read in strong form\footnote{ Equation \eqref{eq:PDE system}$_1$ and \eqref{eq:PDE system}$_2$ together imply $\textrm{Div}\left( \cm\,\sym\,P\right) =f$. Constraining $\sym\,P=\sym\,\nabla u$ gives $\textrm{Div}\left( \cm\,\sym\,\nabla u\right) =f$.}
\begin{align}
 \textrm{Div}\left[ \ce\,\sym\left(\nabla\u-\P\right)+\cc\,\skew\left(\nabla\u-\P\right)\right]\label{eq:PDE system}&=f,\\
 \ce\,\sym\left(\nabla\u-\P\right)+\cc\,\skew\left(\nabla\u-\P\right)-\mathbb{C}_{\textrm{micro}}\,\sym\,\P-\mu\,L_{c}^{2}\,\curl\,\curl\,P&=0.\nonumber 
\end{align}
While the relaxed micromorphic model allows for balance equations in the classical format of the Eringen-Mindlin approach (with second-order force stresses and third order moment stresses), the generalized moment balance \eqref{eq:PDE system}$_2$ is conveniently written in a reduced format with a second-order moment tensor $m=\curl\,P$, as in the better known Cosserat model. In fact, one can say that the relaxed micromorphic model has the full micromorphic kinematics but uses the curvature measure of the Cosserat model. The generalized moment balance \eqref{eq:PDE system}$_2$ can be seen as a tensorial Maxwell-problem due to the $\curl\,\curl$ operation. 

Since only $\sym\,P$ and $\curl\,P$ are controlled in the energy \eqref{eq:bd33} the surprising well-posedness of formulation \eqref{eq:PDE system} has been rigorously shown in appropriate Sobolev spaces crucially based on new coercive Korn-type inequalities for incompatible tensor fields $P$  of the form
\begin{align}\label{Korn}
\exists\,c^+>0:\quad\forall P\in H(\textrm{Curl}),\; P&\times\left. \vec{n}\right|_\Gamma=0\nonumber\\
\left\| \sym\,P\right\|^2_{L^2(\Omega)}&+  \left\| \textrm{Curl}\, P\right\|^2_{L^2(\Omega)}\geq c^+\left( \left\| P\right\|^2_{L^2(\Omega)}+ \left\| \textrm{Curl}\, P\right\|^2_{L^2(\Omega)} \right).
\end{align}
Here $\vec{n}$ is a unit vector orthogonal to $\Gamma\subset\partial\Omega$, see \cite{neff2012maxwell,bauer2016dev,neff2015poincare,neff2006existence,neff2016real}. Note that $P\in H(\curl)$ is not necessarily a continuous field. For $P=\nabla u\in H(\curl)$, inequality \eqref{Korn} turns into a version of Korn's first inequality.

Using \eqref{Korn}, the well-posedness of \eqref{eq:PDE system} then only needs that
$\ce,\cm$ are positive definite, while $\cc$ may be positive semi-definite (e.g., $\cc$ may be even absent) and $\mu\,L_c^2\geq0$, see \cite{neff2015relaxed,neff2014unifying,ghiba2014relaxed}. 

\begin{figure}[H]
	\begin{centering}
		\includegraphics[scale=0.45]{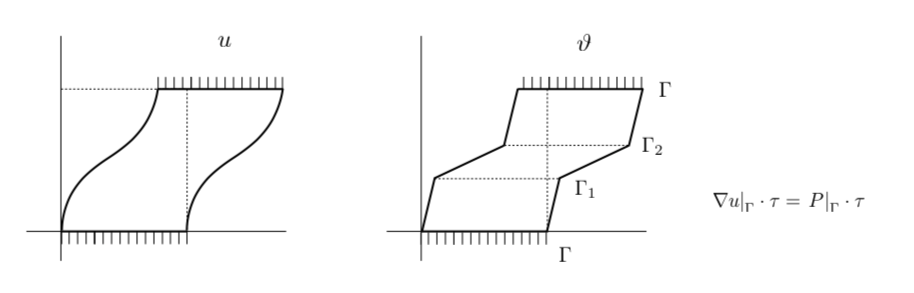}\caption{Shear test on a block. Vertical sides free. Admissible configuration $(u,P)\in H^1(\Omega)\times H(\curl)$ and boundary conditions for the relaxed micromorphic model. Consider $P=\nabla\vartheta\in H(\curl)$, where $\vartheta:\R^3\fr\R^3$ is merely  a Lipschitz function whose weak derivatives $\nabla\vartheta$ may have jumps. Along a given interface $\Gamma_1$, $\Gamma_2$, the micro-distortion $P$ may have normal jumps but is tangentially continuous. The tangential boundary condition for the micro-distortion $P$ applies only at the upper and lower face $\Gamma$ where the displacement $u$ is prescribed.}
		\par\end{centering}
\end{figure}

\subsection{Comparison to Eringen-Mindlin micromorphic models}

The status of the relaxed micromorphic model within the framework of Eringen-Mindlin micromorphic approaches or in relation to higher gradient continua is discussed at length in \cite{neff2014unifying,neff2015relaxed,ghiba2014relaxed,madeo2016complete,madeo2014band}. Here, the following remarks can be made: putting all non-essential differences aside, the energy of a typical Eringen-Mindlin micromorphic approach can be written as 
\begin{equation}\label{EM model}
W_{\textrm{EM}}\left( \nabla u, P, \nabla P\right)=\frac{1}{2}\underbrace{\langle \widehat{\mathbb{C}}_e\left( \nabla u-P\right),\nabla u-P  \rangle_{\R^{3\times 3}}}_{\textrm{relative  elastic energy}}+ \frac{1}{2}\underbrace{\left\langle \cm\,\sym\,P,\sym\,P \right\rangle_{\R^{3\times 3}}}_{\textrm{micro - self - energy}}+\underbrace{\frac{\mu\,L_c^2}{2}\left\|\nabla P \right\|^2}_{\textrm{full curvature}}, 
\end{equation}
where $\widehat{\mathbb{C}}_e$ is a non-standard positive definite fourth-order elasticity tensor, with 45-independent entries, mapping non-symmetric second-order tensors to non-symmetric second-order tensors\footnote{The curvature expression in the Eringen-Mindlin-model or gradient elasticity model would typically include a sixth-order tensor \cite{auffray2009derivation}, in contrast to the relaxed micromorphic model, which only needs a fourth-order tensor.}. Observe that $\nabla P=\nabla e-\nabla\nabla u$ couples derivatives of $e$ and $\nabla u$ contrary to the decoupling in \eqref{curl coupling}. The solution $(u,P)$ will be typically in $H^1(\Omega)\times H^1(\Omega)$.
The standard micromorphic equilibrium equations in strong form read\footnote{\label{EERI} Equation \eqref{Eringen}$_1$ and \eqref{Eringen}$_2$ together imply that $\textrm{Div}\left[ \cm\,\sym\,P-\mu\,L_c^2\Delta P\right] =f$. Here $\cm$ is invariably coupled to the characteristic length $L_c$. Constraining $P=\nabla u$ gives $\textrm{Div}\left[ \cm\,\sym\,\nabla u-\mu\,L_c^2\Delta \nabla u\right] =f$. This is the fourth-order equilibrium equation of the second gradient formulation \eqref{second-gradient}.}
\begin{align}\label{Eringen}
\textrm{Div}\,\widehat{\C}_e\,(\nabla u-P)&=f,\nonumber\\
\widehat{\C}_e\,(\nabla u-P)-\cm\,\sym\,P+\mu\,L_c^2\Delta P&=0.
\end{align}
 The Eringen-Mindlin setting contains gradient elasticity $W_{\textrm{GE}}$ by constraining $\nabla u=P$ (i.e., $\widehat{\mathbb{C}}_e\fr+\infty$), such that the variational problem is based on
\begin{equation}\label{second-gradient}
W_{\textrm{GE}}(\nabla u)=\frac{1}{2}\left\langle \cm\,\sym\,\nabla u, \sym\,\nabla u\right\rangle+\frac{\mu\,L_c^2}{2}\left\| \nabla(\nabla u) \right\|^2.  
\end{equation}
Here, no split in micro- and meso-scale is possible. The solution $u$ will be in $H^2(\Omega)$. On the other hand, $W_{\textrm{EM}}$ can be viewed as a penalty formulation of gradient elasticity (penalty $\ce\fr+\infty$)\footnote{The relaxed micromorphic model cannot be obtained as penalty formulation of gradient elasticity and in a 1-D setting it reduces to linear elasticity with stiffness $\cM$.}.

Next, the Cosserat model does not live on two scales either; it can be seen as the formal singular limit of the relaxed micromorphic model for $\cm\fr+\infty$ in which $P=A$ must be skew-symmetric. The corresponding energy is
\begin{align}\label{Cosserat}
W_{\textrm{Coss}}(\nabla u,A,\curl A)&=\frac{1}{2}\left\langle \ce\,\sym\nabla u,\sym\nabla u \right\rangle + \frac{1}{2}\left\langle \cc\,\skew(\nabla u-A),\skew(\nabla u-A) \right\rangle +\frac{\mu\,L_c^2}{2} \left\| \curl\,A\right\|^2,
\end{align}
with equilibrium equations
\begin{align}
\textrm{Div}\left[ \ce\,\sym\nabla u+\cc\,\skew(\nabla u-A) \right]&=f,\nonumber\\
\skew\left(\cc\,\skew(\nabla u-A)-\mu\,L_c^2\,\curl\,\curl A\right)&=0.
\end{align}
Letting finally $\cc\fr+\infty$, we obtain Toupin's indeterminate couple stress model \cite{munch2017modified,ghiba2016variant,madeo2016new,neff2009subgrid} with energy
\begin{equation}\label{Toupin}
W_{\textrm{Toupin}}(\nabla u,\curl\,\skew\,\nabla u)=\frac{1}{2}\left\langle \ce\,\sym\,\nabla u,\sym\,\nabla u \right\rangle +\frac{ \mu\,L_c^2}{2} \left\| \curl\,\skew\,\nabla u\right\|^2.
\end{equation}
Since $\curl$ is isomorphic to $\nabla$ on skew-symmetric matrices \cite{neff2008curl}, the curvature expression in \eqref{Cosserat} and \eqref{Toupin} is fully general, despite appearance.
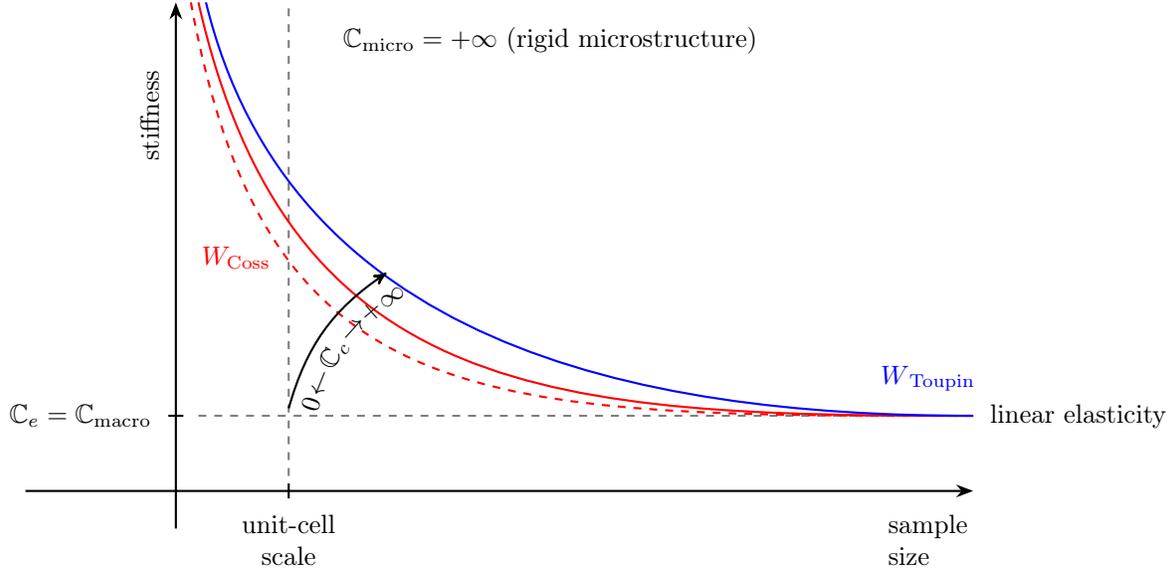
\begin{figure}[H]\label{energies2}
	\begin{centering}
		\begin{tikzpicture}[axis/.style = {help lines,thick,black, -{Stealth[length = 1.5ex]}},brillouin/.style = {domain = -5:10, samples = 100}]\tikzset{%Define style for boxes
			nodeoformula/.style={rectangle,rounded corners=0.2cm,drop shadow={shadow xshift=1mm, shadow yshift=-1mm,opacity=1},draw=black, top color=white, bottom color=white,ultra thick, inner sep=4mm, minimum size=3em, text centered},
			nodepoint/.style={circle,draw=gray,fill=gray,inner sep=0.8mm},
			nodepoint2/.style={circle,very thick,draw=black,inner sep=1.4mm},
			nodepoint3/.style={rectangle, rounded corners, draw=black, very thick, text width=6.5em, minimum height=2em, text centered},
			%Define standard arrow tip
			>=stealth',
			% Define arrow style
			pil/.style={->, thick, shorten <=2pt, shorten >=2pt,}}
		%-------------------Axes--------------------------
		\draw [axis] (-2,-0.5) -- (10.6,-0.5);
		\draw [axis] (0,-1) -- (0,6);
		%-------------------Text--------------------------
		\node[below=2mm,align=left] at (10,-0.5) {sample \\ size};
		\node[] at (12,0.5) {linear elasticity};
		\node[below=2mm,align=center] at (1.5,-0.5) {unit-cell \\ scale};
		\node[left=2mm] at (0,0.5) {$\ce=\cM$};
		\node[rotate=90, above=4mm] at (0.3,4.5) {stiffness};
		\node[right=1mm] at (2,5.5) {$\cm=+\infty$ (rigid microstructure)};
		\node[blue] at (10,1) {$W_{\textrm{\,Toupin}}$};
		\node[red] at (0.8,2.6) {$W_{\textrm{Coss}}$};
		%\node[red] at (0.8,1) {$\cc\fr0$};
		%\node[red] at (2.5,1) {$\mu_c\fr0$};
		%\node[red] at (4,3.2) {$\cc\fr\infty$};
		%\node[red] at (4,2.8) {$\mu_c\fr\infty$};
		%\node[blue] at (10,-0.2) {$L_c\fr0$};
		%\node[blue] at (0.75,-0.2) {$\infty\leftarrow L_c$};
		%-------------------Small lines--------------------------
		\draw [thick] (-0.1,0.5) -- (0.1,0.5);
		%\draw [thick] (-0.1,5.5) -- (0.1,5.5);
		\draw [thick] (1.5,-0.6) -- (1.5,-0.4);
		%-------------------Dotted lines--------------------------
		\draw [thick,dashed,gray] (0.3,0.5) -- (10.6,0.5);
		%\draw [thick,dashed,gray] (0.3,5.5) -- (2,5.5);
		\draw [thick,dashed,gray] (1.5,-0.3) -- (1.5,6);
		%-------------------Plot functions--------------------------
		\draw[thick,red] (0.3,6) .. controls (1.5,0.7) and (5,0.5) .. (10.6,0.5);
		\draw[thick,red,dashed] (0.2,6) .. controls (1.2,0.5) and (4,0.5) .. (10.6,0.5);
		\draw[thick,blue] (0.4,6) .. controls (1.5,1.5) and (6,0.5) .. (10.6,0.5);
		%\draw[ pil, bend left=20, thick, red] (1.5,0.6) to node[auto, swap] {}(2.8,2.4); 
		\def\myshift#1{\raisebox{1ex}}
			%\draw [->,thick,postaction={decorate,decoration={text along path,text align=center,text={|\sffamily\myshift|$ \cc\fr\infty$}}}] (1.5,0.6) to [bend left=20]  (2.8,2.4);
     %\draw [->,thick,postaction={decorate,decoration={text along path,text align=center,text={|\sffamily\myshift|$ \cc\fr\infty$}}}] (1.5,0.6) to [bend left=20]  (2.8,2.4);
     \draw [->,thick,postaction={decorate,decoration={raise=-2.5ex,text along path,text align={center},
     		text={%
     			|\color{black}|
     			 $0{\,}{\leftarrow}{\;}{\mathbb{C}_c}{\;}{\rightarrow}{\,}+{\,}{\infty}$}
 }}] (1.5,0.6) to [bend left=20]  (2.8,2.4);
		\end{tikzpicture}
		\par\end{centering}
	\caption{Qualitative relation between the Cosserat model and Toupin's version of gradient elasticity. Cosserat elasticity can be seen as the formal limit $\cm\fr+\infty$ of the relaxed micromorphic model. Moreover, Cosserat elasticity is a penalty formulation $(\cc\fr+\infty)$ of Toupin's couple-stress model.}
\end{figure}
%0\leftarrow\cc\rightarrow\infty
\subsection{Scaling relations and the relaxed micromorphic model}

\begin{figure}[H]
	\begin{centering}
		\includegraphics[scale=0.23]{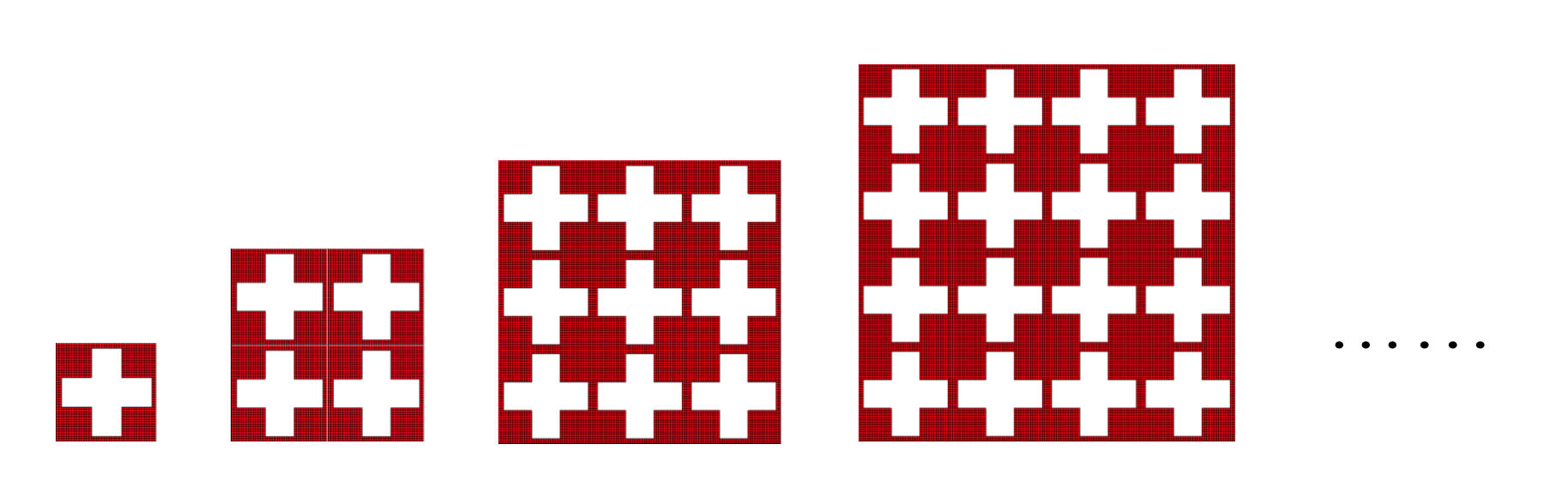}
		\caption{\label{diff}"Smaller is stiffer": $N\times N$ array of unit cells. The apparent stiffness $\C_{\textrm{KUBC}}$ (stiffness under applied kinematically uniform boundary conditions) decreases with increasing size to reach in the limit of infinite size $\cM$. This stiffness-delta between the smallest discernable size and the infinite body will be essential for the determination of the scale-independent parameters in the relaxed micromorphic model and is, on the other hand, responsible for the size-effect. The classical RVE needs to be so large such that its apparent stiffness approximates already that of the macroscopic specimen. Using periodic boundary conditions (PBC) on a unit cell mimics the large scale response and discards the length-scale effect, but allows to determine the effective stiffness on a unit-cell.}
		\par\end{centering}
\end{figure}

The relation of the characteristic length $L_c$ to the dimensions of the material sample follow from a simple scaling argument. Considering a transformation of an arbitrary domain to a unit domain, the length scale $L_c$ w.r.t. the unit domain is inversely proportional to the domain size such that $L_c\fr0$ for very large samples and $L_c\fr\infty$ for small samples. Thus, $L_c$ encodes the relative interaction strength of the microstructure.
It is useful to apply a scaling transformation to the sequence in Figure \ref{diff} and to refer every block of cells to the same unit domain $\left[ 0,1\right]\times\left[ 0,1\right]  $

\begin{figure}[H]
	\begin{centering}
		\includegraphics[scale=0.55]{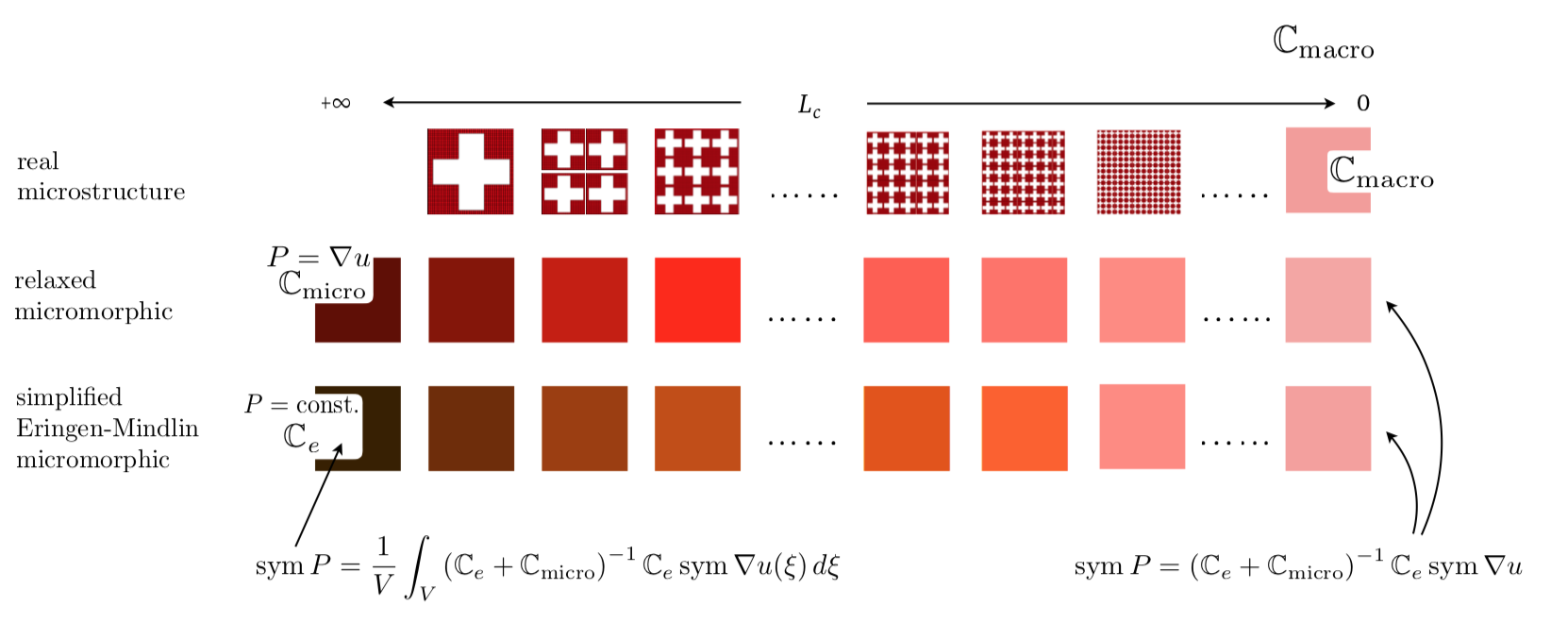}
		\caption{(Stiffness in affine loading w.r.t. displacement $u$, no boundary condition on $P$) First line: $N\times N$ arrays with crosses after transformation to the unit domain. Here, volume fractions of air (white) and aluminium (red) are constant. Note that the formerly smallest block (on the left) still has bounded stiffness under affine Dirichlet boundary conditions. There is no given morphology corresponding to $L_c=+\infty$. For arbitrary many crosses (right) the affine stiffness approximates the effective stiffness $\cM$. Second line: the relaxed micromorphic model is intended to model the response of an $N\times N$ array of crosses, down to perhaps an $3\times3$ or $4\times4$ array and to improve upon a modelling solely with $\cM$ under arbitrary boundary conditions. The characteristic length $L_c$ w.r.t. the unit-domain scales inversely with the size. $L_c\fr\infty$ corresponds to using $P=\nabla u$ and yields linear elasticity with stiffness $\cm$, while letting $L_c\fr0$ yields linear elasticity with stiffness $\cm(\cm+\ce)^{-1}\ce$, which must be identified with $\cM$.  Third line: for comparison, we depicted a simplified Eringen-Mindlin micromorphic model with scale-independent constitutive law coinciding with the relaxed micromorphic model. For $L_c\fr\infty$ the micro-distortion $P$ is necessarily constant and the generated stiffness of the simplified Eringen-Mindlin model in affine loading approximates $\ce$. The constant micro-distortion $P$ gets coupled to a volume average of $\nabla u$. \label{Lc}}
		\par\end{centering}
\end{figure}

The scaling to the unit-domain in the relaxed micromorphic model (Figure \ref{Lc}) modifies only the characteristic length. On the unit-domain we have accordingly

\begin{align}
\int_{\left[ 0,1\right] ^2}\frac{1}{2}\left\langle \mathbb{C}_{e}\,\sym\left(\grad\u-P\right),\sym\left(\grad\u-P\right)\right\rangle&+\frac{1}{2}\left\langle \cc\,\skew\left(\grad\u-P\right),\skew\left(\grad\u-P\right)\right\rangle \nonumber\\
&+\frac{1}{2}\left\langle \mathbb{C}_{\textrm{micro}}\,\sym\,P,\sym\,P\right\rangle +\frac{\mu}{2}\left(\frac{\widehat{L}_c}{N} \right)^{\!\!2} \left\|\curl\,P\right\|^2\,dx\longrightarrow\min.
\end{align}

A very large sample with many crosses $N\fr\infty$ corresponds to $L_c=\frac{\widehat{L}_c}{N}\fr0$ and should deliver linear elasticity with $\cM$ as response of the relaxed micromorphic model. This will generate a first necessary condition between $\ce$, $\cc$, $\cm$ and $\cM$ for a given microstructure, considered in the next subsection.

\subsection{Linear elastic effective macroscopic response and tensor harmonic mean}

\noindent For characteristic length $L_c\fr0$, the relaxed micromorphic equilibrium equations \eqref{eq:PDE system} turn into
\begin{align}\label{Lc_zero}
\textrm{Div}\left[\ce\,\sym\left(\nabla\u-\P\right)+\cc\,\skew\left(\nabla\u-\P\right)\right]&=f,\\
\ce\,\sym\left(\nabla\u-\P\right)+\cc\,\skew\left(\nabla\u-\P\right)-\mathbb{C}_{\textrm{micro}}\,\sym\,\P&=0.\nonumber 
\end{align}
Observing that \eqref{Lc_zero}$_2$ implies $\cc\,\skew\left(\nabla\u-\P\right)=0$ we may solve for $\sym\,P$ in \eqref{Lc_zero}$_2$ and reinserting the result in \eqref{Lc_zero}$_1$ we obtain \cite{barbagallo2016transparent} the \textsl{scale-independent large scale linear elastic response}\footnote{Letting $L_c\fr0$ in \eqref{Eringen} leads to the algebraic side condition $\widehat{\C}_e\,(\nabla u-P)=\cm\,\sym\,P$. Due to the more general format of $\widehat{\C}_e$ as compared to $\ce$ and $\cc$ in \eqref{Lc_zero}, it is not possible to analytically solve for $\sym\,P$ and no transparent formula connecting $\widehat{\C}_e$ and $\cm$ to $\cM$ like \eqref{eq:homog formula} results. The formally scale-independent material parameters of the classical Eringen-Mindlin-model are $\widehat{\C}_e$ and $\cm$ and the scale-independent parameters of $W_{\textrm{GE}}$ are $\cm=\cM$. For the Cosserat model, the respective scale-independent stiffness is $\ce=\cM$. However, considering (footnote \ref{EERI}) $\textrm{Div}\left[ \cm\,\sym\,P-\mu\,L_c^2\Delta P\right] =f$, it is not strictly possible to say that $\cm$ is scale-independent in the Eringen-Mindlin model. The identification of $\cm$ (and therefore also $\widehat{\C}_e$) in the Eringen-Mindlin model may be length-scale dependent after all. }
\begin{equation}
\textrm{Div}\,\cM\,\sym\,\nabla\u=f.
\end{equation}
In \cite{barbagallo2016transparent} it is shown that the effective macroscopic elasticity tensor $\cM$ is exactly given by the Reuss lower bound \cite{lobos2017homogenization,reuss1929berechnung} of the meso-scale stiffness $\ce$ and the micro-scale stiffness $\cm$
\begin{align}
\cM&=\mathbb{C}_{\textrm{micro}}\left(\mathbb{C}_{\textrm{micro}}+\mathbb{C}_{e}\right)^{-1}\mathbb{C}_{e}=\left(\mathbb{C}_{e}^{-1}+\mathbb{C}_{\textrm{micro}}^{-1}\right)^{-1},&\mathbb{C}_{e}&=\mathbb{C}_{\textrm{micro}}\left(\mathbb{C}_{\textrm{micro}}-\mathbb{C}_{\textrm{macro}}\right)^{-1}\mathbb{C}_{\textrm{macro}}.\label{eq:homog formula}
\end{align} 
Remark that $\cc$ does not intervene in \eqref{eq:homog formula} and therefore the scale-independent material parameters in the relaxed micromorphic model are only $\ce$ and $\cm$. 
If we specialize formula \eqref{eq:homog formula} to the isotropic relaxed micromorphic model, the
macroscopic coefficients of the equivalent macroscopic continuum are related to
the parameters of the relaxed micromorphic model through the homogenization
formulas (see \cite{neff2007geometrically,neff2004material,barbagallo2016transparent})
\begin{equation}
\mu_{\textrm{macro}}=\frac{\me\,\mu_{\textrm{micro}}}{\me+\mu_{\textrm{micro}}},\qquad 2\mu_{\textrm{macro}}+3\lambda_{\textrm{macro}}=\frac{\left(2\me+3\le\right)\left(2\mu_{\textrm{micro}}+3\lambda_{\textrm{micro}}\right)}{\left( 2\me+3\le\right) +\left( 2\mu_{\textrm{micro}}+3\lambda_{\textrm{micro}}\right)}.\label{neff}
\end{equation}

These formulas are each identical to calculating the equivalent stiffness of two linear springs in series: for (\ref{neff})$_{1}$ it is the equivalent stiffness $\mu_\textrm{macro}$ from the springs with stiffness $\me$ and $\mh$, for (\ref{neff})$_{2}$ the equivalent stiffness $2\mu_{\textrm{macro}}+3\lambda_{\textrm{macro}}$  is analogously obtained from the springs in series of stiffness $2\mu_{e}+3\lambda_{e}$ and of stiffness $2\mu_{\textrm{micro}}+3\lambda_{\textrm{micro}}$. Such formulas, as we will show in this paper
for the anisotropic case, are essential to characterize the mechanical
behavior of heterogeneous metamaterials on different scales. 

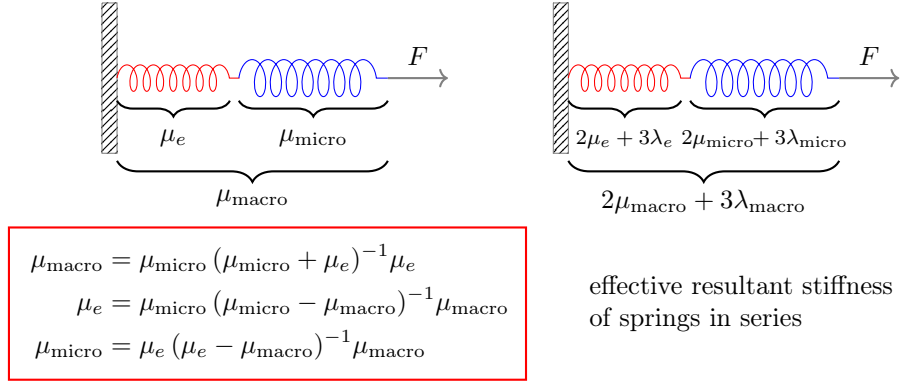
\begin{figure}[H]
	\begin{centering}
		\begin{tikzpicture}
			\node (M1) [circle] at (2,-1) {};
		%------------------------------------------------------------------------
		\filldraw[gray,pattern=north east lines] (0,0) -| (0.2,-2) -|  (0,0);
		\draw [ snake=coil, red, segment amplitude=5pt, segment length=5pt] (0.2,-1) -- (M1);
		\draw [ snake=coil, blue, segment amplitude=7pt, segment length=6pt] (1.82,-1) -- (3.8,-1);
		%------------------------------------------------------------------------
		\draw [thick,decoration={brace,amplitude=6pt,mirror,raise=0.3cm},decorate] (0.2,-1) -- (1.7,-1) 
		node [pos=0.5,anchor=north,yshift=-0.55cm] {$\me$}; 
		\draw [thick,decoration={brace,amplitude=6pt,mirror,raise=0.3cm},decorate] (1.82,-1) -- (3.8,-1) 
		node [pos=0.5,anchor=north,yshift=-0.55cm] {$\mh$}; 
		\draw [thick,decoration={brace,amplitude=8pt,mirror,raise=0.3cm},decorate] (0.2,-1.8) -- (3.8,-1.8) 
		node [pos=0.5,anchor=north,yshift=-0.55cm] {$\mum$}; 
		%------------------------------------------------------------------------
		%------------------------------------------------------------------------
		\node (M2) [circle,xshift=6cm] at (2,-1) {};
		%------------------------------------------------------------------------
		\filldraw[gray,pattern=north east lines,xshift=6cm] (0,0) -| (0.2,-2) -|  (0,0);
		\draw [ snake=coil, red, segment amplitude=5pt, segment length=5pt,xshift=6cm] (0.2,-1) -- (M2);
		\draw [ snake=coil, blue, segment amplitude=7pt, segment length=6pt,xshift=6cm] (1.82,-1) -- (3.8,-1);
		%------------------------------------------------------------------------
		\draw [thick,decoration={brace,amplitude=6pt,mirror,raise=0.3cm},decorate,xshift=6cm] (0.2,-1) -- (1.7,-1) 
		node [pos=0.5,anchor=north,yshift=-0.55cm] {{\footnotesize $2\mu_{e}+3\lambda_{e}$}}; 
		\draw [thick,decoration={brace,amplitude=6pt,mirror,raise=0.3cm},decorate,xshift=6cm] (1.82,-1) -- (3.8,-1) 
		node [pos=0.5,anchor=north,yshift=-0.55cm] {{\footnotesize $2\mu_{\textrm{micro}}\!\!+3\lambda_{\textrm{micro}}$}}; 
		\draw [thick,decoration={brace,amplitude=8pt,mirror,raise=0.3cm},decorate,xshift=6cm] (0.2,-1.8) -- (3.8,-1.8) 
		node [pos=0.5,anchor=north,yshift=-0.55cm] {$2\mu_{\textrm{macro}}+3\lambda_{\textrm{macro}}$}; 
		\node[align=left] at (8.5,-4) {effective resultant stiffness \\  of springs in series};
		\node[draw=red,align=left,inner sep=1.5ex,thick] at (2.2,-4) {$\begin{aligned}
			\mum&=\mh\,(\mh+\mu_{e})^{-1}\mu_{e} \\ 
			\mu_{e}&=\mu_{\textrm{micro}}\,(\mu_{\textrm{micro}}-\mu_{\textrm{macro}})^{-1}\mu_{\textrm{macro}}\\
			\mu_{\textrm{micro}}&=\mu_{e}\,(\mu_{e}-\mu_{\textrm{macro}})^{-1}\mu_{\textrm{macro}}
			\end{aligned}$} ;
		\draw[->, thick,gray] (3.8,-1) -- (4.6,-1);
		\draw[->, thick,gray] (9.8,-1) -- (10.6,-1);
		\node[] at (4.2,-0.7) {$F$};
		\node[] at (10.2,-0.7) {$F$};
		\end{tikzpicture}
		\par\end{centering}
	\caption{Scale-independent response governed by two springs in series. If $\mh=\mum$ then $\me=+\infty$.}
\end{figure}

As it will turn out, the modeling perspective of the relaxed micromorphic model endows the macro-scale as well as the micro-scale with sufficient physical characteristics to define the setting how to compute the parameters of each of these scales by first order homogenization. Put different and with reference to the stiffness laws for springs in series, the equivalent stiffness and the stiffness of one spring in series can be identified. The unknown stiffness of the second spring representing the transition scale however, is obtained from (\ref{neff})$_{1}$ and (\ref{neff})$_{2}$ by solving for $\mu_e$ and $\lambda_e$.

\subsection{Microscopic response - zoom into the microstructure}

\begin{figure}[H]
	\begin{centering}
		\begin{tikzpicture}[axis/.style = {help lines,thick,black, -{Stealth[length = 1.5ex]}},brillouin/.style = {domain = -5:10, samples = 100}]
		\tikzset{>=stealth',
			% Define arrow style
			pil/.style={->, thick, shorten <=2pt, shorten >=2pt,}}
		%-------------------Axes--------------------------
		\draw [axis] (-2,0) -- (10.6,0);
		\draw [axis] (0,-1) -- (0,7);
		%-------------------Text--------------------------
		\node[below=2mm,align=left] at (10,0) {sample \\ size};
		%\node[left=5mm,rotate=90] at (0,5.8) {};
		\node[left=20mm,rotate=90] at (0,5.8) {$\leftarrow$ "relax" stiffness};
		\node[below=2mm,align=center] at (1.5,0) {unit-cell \\ scale};
		\node[] at (11.97,3) {linear elasticity with $\cm$};
		\node[] at (12,1.4) {linear elasticity with $\cM$};
		\node[left=1.4mm] at (0,1) {$\cM$};
		%\node[left=2mm] at (0,0.5) {$\ce$};
		%\node[left=2mm] at (0,0) {$\displaystyle\frac{1}{2}\;\cM$};
		\node[left=2mm] at (0,2.5) {$\cm$};
		%\node[left=2.2mm] at (0,4.5) {$\widehat{\C}_e$};
		\node[] at (-0.82,3) {$\C_{\textrm{Voigt}}$};
		\node[left=2mm,rotate=90] at (-0.28,2.2) {$\textcolor{red}{-\!\!\!-\!\!\!\longrightarrow}$};
		\node[left=2mm] at (-0.35,1.72) {$\textcolor{red}{\mathbb{C}_{e}\fr+\infty}$};
		\node[left=2mm] at (6.4,2.2) {$\textcolor{blue}{\widehat{\C}_e\fr+\infty}$};
		\node[red] at (7,3) {$W_{\textrm{GE}}$};
		\node[red] at (0.5,2.8) {$W$};
		\node[blue] at (7,1.4) {$W_{\textrm{EM}}$};
		%\node[blue] at (10,0.3) {$L_c\fr0$};
		%\node[blue] at (0.75,0.3) {$\infty\leftarrow L_c$};
		%-------------------Small lines--------------------------
		\draw [thick] (-0.1,1) -- (0.1,1);
		%\draw [thick] (-0.1,0) -- (0.1,0);
		\draw [thick] (-0.1,2.5) -- (0.1,2.5);
		\draw [thick] (1.5,-0.1) -- (1.5,0.1);
        \draw [thick] (-0.1,3) -- (0.1,3);
		%\draw [thick] (-0.1,-1) -- (0.1,-1);
		%\draw [thick] (-0.1,0.7) -- (0.1,0.7);
		%-------------------Dotted lines--------------------------
		\draw [thick,dashed,gray] (0.3,1) -- (10.6,1);
		%\draw [thick,dashed,gray] (0.3,0) -- (10.6,0);
		%\draw [thick,dashed,gray] (0.3,0.7) -- (10.6,0.7);
		\draw [thick,dashed,gray] (0.3,2.5) -- (10.6,2.5);
		\draw [thick,dashed,gray] (1.5,0.2) -- (1.5,6);
		%-------------------Plot functions--------------------------
		\draw[thick,red] (1,7) .. controls (1.5,4) and (5,2.6) .. (7.6,2.6);
		\draw[thick,blue,dashed] (0.5,7) .. controls (0.9,3.8) and (3,1.05) .. (7.6,1.05);
		\draw[thick,blue] (0.7,7) .. controls (1.2,4.2) and (3,1.05) .. (7.6,1.05);
		\draw[thick,red] (0.1,2.5) .. controls (3.4,2.6) and (2.5,0.9)  .. (7.6,1.02);
		\draw[thick,red,dashed] (0.1,2.5) .. controls (3.4,2.0) and (2.5,0.9)  .. (7.6,1.02);
		\draw[thick,red,dashed] (0.1,2.5) .. controls (3.4,1.4) and (2.5,0.9)  .. (7.6,1.02);
		\draw[thick,red,dashed] (0.1,2.5) .. controls (2.4,1) and (2.5,0.9)  .. (7.6,1.02);
		\draw [thick,blue] (7.6,1.05) -- (10.6,1.05);
		\draw [thick,red] (7.6,1.02) -- (10.6,1.02);
		\draw [thick,red] (7.6,2.6) -- (10.6,2.6);
		%-------------------Arrows--------------------------
		\node[rotate=90,fill=white, inner sep=1mm] at (1.1,1.8) {$ 0\leftarrow L_c\fr\infty$};
		\draw[ pil, bend left=20, thick, blue,xshift=3mm,yshift=2mm] (3.8,1.6) to node[auto, swap] {}(4.6,2.8); 
		%\draw [->,very thick,black] (1.3,1.5) -- node[rotate=90,above=1mm,fill=white, inner sep=1mm] {$ L_c\fr0$} (1.3,0.7);
		\end{tikzpicture}
		\par\end{centering}
	\caption{(Stiffness in non-affine loading, like torsion) Qualitative difference of the relaxed micromorphic model $W$, the standard Eringen-Mindlin model $W_{\textrm{EM}}$ and gradient elasticity $W_{\textrm{GE}}$. In all three cases "smaller is stiffer". Typically, the Eringen-Mindlin model and gradient elasticity model dramatically overestimate the stiffness and both show unbounded stiffness for small sample sizes. Only the relaxed micromorphic model has bounded stiffness for all sample sizes (including below a unit-cell level) as well as a clear separation of microscopic and macroscopic scales, $\cm$ and $\cM$, respectively. For a microscopically homogeneous material, $\cm=\cM$ induces $\ce=+\infty$ which enforces $\sym\,P=\sym\,\nabla u$ and yields linear elastic response with $\C_{\textrm{macro}}$ for the relaxed micromorphic model, while the constraint $\widehat{\C}_e=+\infty$ enforces $P=\nabla u$ in $W_{\textrm{EM}}$ and generates the second gradient model $W_{\textrm{GE}}$ with unbounded stiffness. The stiffness of the relaxed micromorphic model $W$ is "relaxed" as compared to $W_{\textrm{EM}}$. Note that the characteristic length scale $L_c>0$ always interferes with the actual stiffness at the unit-cell level.}\label{energies}
\end{figure}
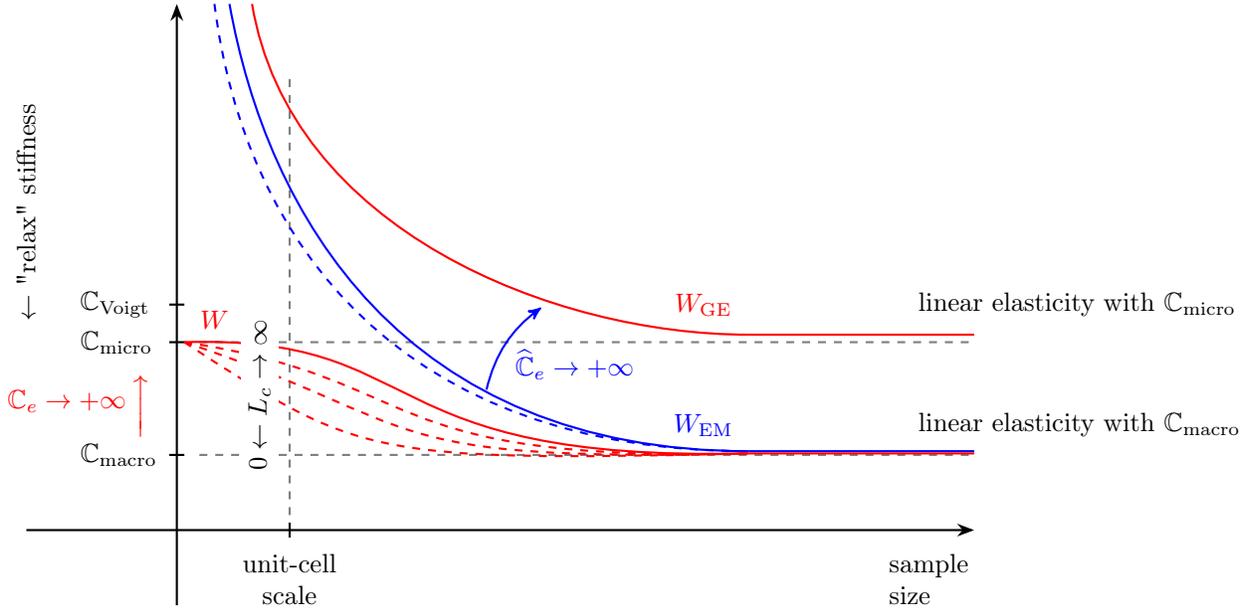

Next, we consider the limit $L_c\fr\infty$ for the characteristic length (see Figure \ref{Lc} and Figure \ref{energies}). Looking at the variational formulation \eqref{ener1},\eqref{eq:bd33}, we see that in simply connected domains $\Omega$ this generates the constraint $P=\nabla\vartheta$ for some function $\vartheta:\Omega\subseteq\R^3\fr\R^3$ due to the precence of the $\curl-$curvature measure. Together with the appropriate tangential boundary condition \cite{neff2014unifying,ghiba2014relaxed} $\left.P \right|_{\partial\Omega}\cdot\tau=\left.\nabla u \right|_{\partial\Omega}\cdot\tau $ we obtain the new minimization problem
\begin{equation}
\int_\Omega W(\nabla u,\nabla\vartheta,0)\,dx\;\longrightarrow\;\min.\qquad(u,\vartheta)\in H^1(\Omega)\times H^1(\Omega),
\end{equation}
the solution of which necessitates $u=\vartheta$ and the remaining macroscopic displacement field $u$ is obtained from the minimization problem
\begin{equation}
\int_\Omega \frac{1}{2} \left\langle \cm\,\sym\,\nabla u,\sym\,\nabla u \right\rangle \,dx\;\longrightarrow\;\min.\qquad u\in H^1(\Omega).
\end{equation}
This reduction feature depends critically on using $\curl\,P$ as curvature measure and clearly distinguishes the relaxed micromorphic model from the standard Eringen-Mindlin approaches which use $\nabla P$ as curvature measure.

\medskip

The main task which has to be accomplished to successfully apply constitutive
laws to the real material world is the identification of their parameters.
For the present anisotropic relaxed micromorphic model there are essentially three sets of
material parameters $\cm$, $\mathbb{C}_{e}$ and $\cM$ describing
the scale-independent static response. By virtue of the recently established
harmonic mean-type micro-macro homogenization rule (\ref{eq:homog formula}),
the elasticities of the three scales are connected such that the determination
of two sets is enough to infer on the third one. Only the micro-scale $\cm$
as well as mesoscale-elastic parameters $\ce$ appear directly in
the relaxed micromorphic energy (\ref{eq:bd33}), while $\cM$
refers to a macroscopic, energetically equivalent, linear elastic
surrogate model. In view of these characteristics\footnote{It is indeed well known in the field of homogenization techniques
	(see, e.g., \cite{Eidel,pecullan1999scale}) that the homogenization
	of a unit-cell on which one imposes periodic boundary conditions mimics
	the behavior of a very large specimen of the associated equivalent
	Cauchy continuum. Usually, homogenization techniques only provide
	a direct transition from the micro to the macro-scale without considering
	the intermediate (transition) scale in which all relevant microstructure-related
	phenomena are manifest. Some attempts to introduce a transition scale
	via the homogenization towards a micromorphic continuum are made in
	\cite{hutter2017homogenization,trinh2012evaluation}, even if it is
	clear that a definitive answer is far from being provided (see \cite{hutter2017homogenization,trinh2012evaluation}
	and references cited there). Our relaxed micromorphic model naturally
	provides the bridge between the micro and macro behavior of the considered
	homogenized material with the simple and transparent tensor homogenization
	formulas \eqref{eq:homog formula}. }, the coefficients
of $\cM$ can be identified by standard homogenization on the periodic
unit-cell level. However, the identification of the micro-scale parameters
$\cm$ through homogenization is completely non-standard, since novel
criteria have to be established to identify from all possible unit-cell variants (notice that for periodicity the unit-cell is unique but its representation is not) those which are consistent with the present modeling. Similarly, the boundary conditions have to be consistently chosen.

\subsection{Outline}

\noindent In this paper, then, we will propose a method to quantify $\cm$ (and a fortiori $\ce$) for a given microstructure. The conceptual idea is simple. Instead of subjecting the relaxed micromorphic model  to large specimen size ($L_c\fr0$), we consider arbitrarily small specimen sizes ($L_c\fr+\infty$). For this formal limit, the relaxed micromorphic model turns as well into a linear elastic boundary value problem with constant stiffness tensor $\cm$. It is important to note at this point that it is meaningless to follow this idea in the context of standard Eringen-Mindlin or higher gradient and Cosserat continua since they have unbounded stiffness for $L_c\fr+\infty$ (see Figure~\ref{energies}). 

Having a precise formal limit for $L_c\fr+\infty$ at hand we realize, that the smallest resolution we may actually consider is the size of a unit-cell of the material which is to be described.  Moreover, using affine Dirichlet boundary conditions, the resulting stiffness depends on the size and shape of the unit-cell where, in general, a smaller unit-cell is "stiffer" than a larger unit-cell. In order to reasonably match the actual response of the given microstructure for small specimen size to $\cm$, we need to observe that i) there are several valid variants of the unit-cell (different sizes and geometries generate the same macroscopic specimens), and ii) among these unit-cells there is in general not a single stiffest one. Therefore, we will be satisfied with a quantitative estimate relating $\cm$ and calculated stiffnesses for several unit-cells under affine Dirichlet boundary conditions.

For consistency in the scale transition process it is a necessity to claim that the micro-scale reflects and preserves the existing material symmetries of the macro-scale. For that reason we have to invoke Neumann's principle on the representation of anisotropy. As a consequence of this requirement the number of candidate unit-cells is drastically reduced.

We will be left with an estimate of the sort that $\cm$ of the relaxed micromorphic model should bound all obtained stiffnesses of competing unit-cells in the energy norm.
Insisting on an optimal choice for $\cm$ we try to determine one stiffness tensor, which is the least stiff, satisfying the foregoing estimates. This leads us formally to the so called Löwner matrix supremum problem, the solution of which is in general not unique. In the application to tetragonal planar metamaterials we are considering, however, it will be shown that this tensor is uniquely determined.

The only short-range elastic parameter which is not yet determined by the presented
arguments is the generalized Cosserat couple modulus $\cc$, but it can be
evaluated when considering the dynamical analysis of the proposed
metamaterial \cite{d2019effective}.

\section{\label{subsec:Mathematical-arguments-concernin}Rigorous determination
	of $\mathbb{C}_{\textrm{micro}}$}

\subsection{Maximal stiffness on the micro-scale}

In this section we describe the mathematical underpinning towards
determining the stiffness $\cm$ in the relaxed micromorphic model.
We do this in the static case, in which the equilibrium problem (\ref{eq:PDE system})
can be obtained as the energy minimization problem
\begin{equation}
I\left(u,P\right):=\int_{\Omega}W\left(\nabla u,P,\textrm{Curl}\,P\right)dx\longrightarrow\min.\quad\left(u,P\right)\in H^{1}\!\left(\Omega\right)\times H\!\left(\curl\right),\label{Gamma problem}
\end{equation}
under suitable boundary conditions, with $W$ given in \eqref{eq:bd33}.
For the displacement field $u$ we apply overall affine Dirichlet
boundary conditions
\begin{equation}
\left.u\right|_{\partial\Omega}\left(x\right)=\overline{B}\cdot x,\qquad\overline{B}\in\R^{3\times3}\label{bd1}
\end{equation}
and the micro-distortion tensor $P$ has to satisfy the compatible boundary
condition\footnote{In this way, artificial boundary layer effects are avoided.}
\begin{equation}
\left.\nabla u\right|_{\partial\Omega}\left(x\right)\cdot\tau_{1,2}=\left.P\right|_{\partial\Omega}\left(x\right)\cdot\tau_{1,2}\,,\label{bd2}
\end{equation}
where $\tau_{1,2}$ are linear independent tangent vectors to $\partial\Omega$. One
then observes that the minimal energy content of a solution $\left(u,P\right)$
to the minimization problem (\ref{Gamma problem}), \eqref{eq:bd33},
(\ref{bd1}), (\ref{bd2}) is easily bounded above by choosing the
macroscopic fields $\left(u,P \right) $ such that 
\begin{equation}
\nabla u\left(x\right)=P\left(x\right),\qquad x\in\Omega.
\end{equation}
This gives the estimate
\begin{equation}
\inf_{\left(u,P\right)}\int_{x\in\Omega}W\left(\nabla u,P,\textrm{Curl}\,P\right)dx\leq\inf_{u}\int_{x\in\Omega}W\left(\nabla u,\nabla u,0\right)dx=\inf_{u}\int_{x\in\Omega}\frac{1}{2}\left\langle \cm\,\sym\,\nabla u\left(x\right),\sym\,\nabla u\left(x\right)\right\rangle dx.\label{eq:inf uP}
\end{equation}
Therefore, the maximal possible stored elastic energy of the relaxed
micromorphic model over an arbitrary window $\widetilde{\Omega}\subset\Omega$
is
\begin{equation}
\inf_{u}\int_{x\in\widetilde{\Omega}}\frac{1}{2}\left\langle \cm\,\sym\,\nabla u\left(x\right),\sym\,\nabla u\left(x\right)\right\rangle dx,\qquad\left.u\right|_{\partial\widetilde{\Omega}}\left(x\right)=\overline{B}\cdot x,\label{eq:integr}
\end{equation}
and this value is attained for $\nabla u\left(x\right)=P\left(x\right)$
for all $x\in\Omega$. Below, we will evaluate the latter condition
over a given unit-cell $V\!\left(x\right)=\widetilde{\Omega}$ attached
at the macroscopic point $x\in\Omega$. Following classical analysis, the average displacement gradient over the unit-cell satisfies \cite[3.1, $u\in C^{ \infty}$]{zohdi2004homogenization}
\begin{align}
\frac{1}{\left|V\!\left(x\right)\right|}\int_{\xi\in V\!\left(x\right)}\nabla_{\xi}u\left(\xi\right)d\xi & =\frac{1}{\left|V\!\left(x\right)\right|}\int_{\xi\in\partial V\!\left(x\right)}u\left(\xi\right)\otimes n\left(\xi\right)\,dS= \frac{1}{\left|V\!\left(x\right)\right|}\int_{\xi\in\partial V\!\left(x\right)}\left(\overline{B}\cdot\xi\right)\otimes n\left(\xi\right)\,dS\nonumber\\
&=\frac{1}{\left|V\!\left(x\right)\right|}\int_{\xi\in V\!\left(x\right)}\nabla_\xi\left[ \overline{B}\cdot\xi\right]  d\xi=\frac{1}{\left|V\!\left(x\right)\right|}\int_{\xi\in V\!\left(x\right)}\overline{B}\,d\xi=\overline{B}.\label{eq:37} 
\end{align}
Symmetrization yields as well for the averaged strain tensor
\begin{equation}\label{epsbar}
\overline{\varepsilon}\defi\frac{1}{\left|V(x)\right|}\int_{\xi\in V\left(x\right)}\sym\nabla_\xi u\left(\xi\right)d\xi=\sym\,\overline{B}=\overline{E}.
\end{equation}
Since the integrand in (\ref{eq:integr}) is convex (quasiconvex)
and $\cm$ is constant by assumption, the maximal storage of elastic
energy in $V\!\left(x\right)$, according to the relaxed micromorphic
model is realized already by the homogeneous displacement $\overline{B}\cdot\xi$
which yields
\begin{align}
\inf\left\{ \int_{\xi\in V\left(x\right)}\frac{1}{2}\left\langle \cm\,\sym\,\nabla_{\xi}\widetilde{v}\left(\xi\right),\sym\,\nabla_{\xi}\widetilde{v}\left(\xi\right)\right\rangle d\xi\;\Bigr|\;\widetilde{v}:V\!\left(x\right)\fr\R^{3},\;\left.\widetilde{v}\right|_{\partial V\left(x\right)}\left(\xi\right)=\overline{B}\cdot\xi\right\} \qquad\qquad\label{Hills}\\
=\frac{1}{2}\left\langle \cm\,\sym\,\overline{B},\sym\,\overline{B}\right\rangle \left|V\!\left(x\right)\right|=\frac{1}{2}\left\langle \cm\,\overline{E},\overline{E}\right\rangle \left|V\!\left(x\right)\right|,\qquad\overline{E}=\sym\,\overline{B},\nonumber 
\end{align}
for the relaxed micromorphic model. 

\bigskip

Now we switch to considering the unit-cell as described by classical
linear elasticity with inhomogeneous material properties. Macroscopic variables are conceptionally some ``averages''
over the micro-scale. Hence the attached unit-cell $V\!\left(x\right)$
must be considered to be loaded such that it produces the given superposed
macroscopic average $\nabla u\left(x\right)$. There are several choices
satisfying this requirement; prominently affine Dirichlet boundary conditions (or KUBC - kinematically uniform boundary conditions) or periodic boundary conditions (PBC). It is well
known that affine Dirichlet conditions generate stiffer response than
PBC \cite{huet1990application,pecullan1999scale,kanit2003determination}. Let $\mathbb{C}(\xi)$ be the geometry dependent inhomogeneous elasticity tensor of the given metamaterial. Under affine Dirichlet
conditions (KUBC) the classical linear elastic stored energy of the
unit-cell is given by
\begin{gather}
\inf\left\{ \int_{\xi\in V\left(x\right)}\frac{1}{2}\left\langle \mathbb{C}\left(\xi\right)\sym\nabla_{\xi}\widetilde{v}\left(\xi\right),\sym\nabla_{\xi}\widetilde{v}\left(\xi\right)\right\rangle d\xi\:\Bigl|\:\widetilde{v}:V\!\left(x\right)\fr\R^{3},\;\left.\widetilde{v}\right|_{\partial V\!\left(x\right)}\left(\xi\right)=\overline{B}\cdot\xi\right\} \qquad\qquad\qquad\qquad\qquad\qquad\qquad\qquad\nonumber \\
=\inf\left\{ \int_{\xi\in V\left(x\right)}\frac{1}{2}\left\langle \mathbb{C}\left(\xi\right)\sym\left(\nabla_{\xi}\left[v\left(\xi\right)+\overline{B}\cdot\xi\right]\right),\sym\left(\nabla_{\xi}\left[v\left(\xi\right)+\overline{B}\cdot\xi\right]\right)\right\rangle d\xi\:\Bigl|\:v\in C_{0}^{\infty}\left(V\!\left(x\right),\R^{3}\right)\right\} \label{fine energy}\\
=\inf\left\{ \int_{\xi\in V\left(x\right)}\frac{1}{2}\left\langle \mathbb{C}\left(\xi\right)\left(\sym\nabla_{\xi}v\left(\xi\right)+\overline{E}\right),\sym\,\nabla_{\xi}v\left(\xi\right)+\overline{E}\right\rangle d\xi\:\Bigl|\:v\in C_{0}^{\infty}\left(V\!\left(x\right),\R^{3}\right)\right\} \nonumber \\
\!\!\!\!\!\!\!\!\!\!\!\!\!\!\!\!\!\!\!\!\!=\int_{\xi\in V\left(x\right)}\frac{1}{2}\left\langle \mathbb{C}\left(\xi\right)\left(\sym\,\nabla_{\xi}\widehat{v}_{\,\overline{E}}\left(\xi\right)+\overline{E}\right),\sym\,\nabla_{\xi}\widehat{v}_{\,\overline{E}}\left(\xi\right)+\overline{E}\right\rangle d\xi,\nonumber 
\end{gather}
$\textrm{where (the corrector) }\widehat{v}_{\,\overline{E}}\in C_{0}^{\infty}\left(V\!\left(x\right),\R^{3}\right)\textrm{ is the realizing minimizer}$
(which is not known a priori).

In the context of homogenization, we now demand that the (fully resolved
linear elastic) fine-scale energy (\ref{fine energy}) should equal
the (relaxed micromorphic) coarse-scale energy (\ref{eq:inf uP})$_{\textrm{left}}$
over the same domain $V\!\left(x\right)$, under the same affine Dirichlet-boundary
conditions $\overline{B}\cdot\xi$ and for the same material. This
means we require that for all affine loadings $\overline{B}\in\R^{3\times3}$ we have:

\begin{gather}
\inf_{\left(u,P\right)}\left\{ \int_{\xi\in V\left(x\right)}W\left(\nabla_{\xi}u\left(\xi\right),P\left(\xi\right),\curl\,P\left(\xi\right)\right)d\xi\,\Bigr|\,\left.u\right|_{\xi\in\partial V}\left(\xi\right)=\overline{B}\cdot\xi,\;\overline{B}\cdot\tau_{1,2}=\left.P\right|_{\xi\in\partial V}\left(\xi\right)\cdot\tau_{1,2}\right\} \nonumber \\
\overset{!}{=}\inf_{v}\left\{ \int_{\xi\in V\left(x\right)}\frac{1}{2}\left\langle \mathbb{C}\left(\xi\right)\left(\sym\nabla_{\xi}v+\overline{E}\right),\sym\nabla_{\xi}v+\overline{E}\right\rangle d\xi\,\Bigr|\,v\in C_{0}^{\infty}\left(V\!\left(x\right),\R^{3}\right),\;\overline{E}=\sym\,\overline{B}\right\} .\label{eq:euqality energies}
\end{gather}
\begin{figure}[H]
	\begin{centering}
		\includegraphics[scale=0.35]{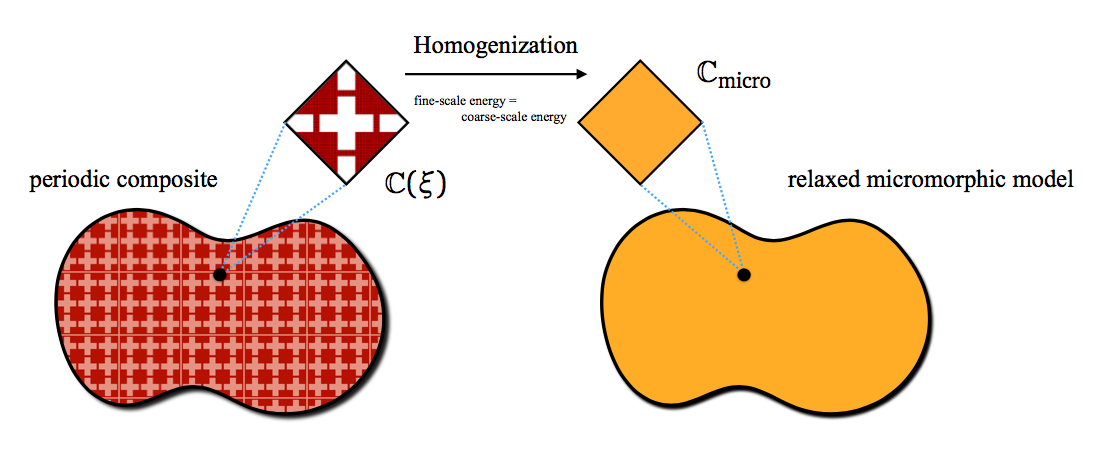}\caption{The process of homogenization. Following the classical Hill-Mandel lemma, we demand energy equivalence
			of the fine-scale linear elastic energy (left) with the coarse-scale
			relaxed micromorphic energy (right) over the same domain $V\!\left(x\right)$,
			under the same affine Dirichlet-boundary conditions and for the same material
			that is to be described.}
		\par\end{centering}
\end{figure}
With estimate (\ref{eq:inf uP}) and (\ref{Hills}), taking (\ref{eq:euqality energies})
into account, we obtain the inequality
\begin{equation}
\underbrace{\frac{1}{2}\left\langle \cm\,\overline{E},\overline{E}\right\rangle \left|V\!\left(x\right)\right|}_{\overset{{\scriptstyle \textrm{coarse scale micromorphic }}}{\textrm{upper energy bound}}}\geq\underbrace{\inf\biggl\{\int_{\xi\in V\left(x\right)}\frac{1}{2}\left\langle \mathbb{C}\left(\xi\right)\left(\sym\nabla_{\xi}v\left(\xi\right)+\overline{E}\right),\sym\nabla_{\xi}v\left(\xi\right)+\overline{E}\right\rangle d\xi\:\Bigl|\:v\in C_{0}^{\infty}\left(V\!\left(x\right),\R^{3}\right)\biggr\}}_{\textrm{fine-scale linear elastic energy}},\label{uguaglianza energie}
\end{equation}
which must be satisfied for all $\overline{E}\in\Sym$.
On the other hand, according to the classical\footnote{And not any of the ambiguous extended versions for generalized continua
	\cite{forest2011generalized,forest2002homogenization,forest1999aufbau,hutter2019micro}.} Hill-Mandel
 lemma \cite{hill1972constitutive,nemat2013micromechanics,hill1963elastic,mandel1971plasticite,zohdi2004homogenization}
we can define a unique apparent \cite{huet1990application} constant stiffness
tensor $\mathbb{C}_{\textrm{KUBC}}^{V}$, independent of $\overline{E}$,
but depending on the chosen unit-cell $V,$ by setting\footnote{Since 
\begin{align}
\frac{1}{2}\left\langle \mathbb{C}_{\textrm{KUBC}}^{V}\,\overline{E},\overline{E}\right\rangle\left|V\left(x\right)\right| &=\inf\biggl\{\int_{\xi\in V\left(x\right)}\frac{1}{2}\left\langle \mathbb{C}\left(\xi\right)\left(\sym\nabla_{\xi}v\left(\xi\right)+\overline{E}\right),\sym\nabla_{\xi}v\left(\xi\right)+\overline{E}\right\rangle d\xi\:\Bigl|\:v\in C_{0}^{\infty}\left(V\!\left(x\right),\R^{3}\right)\biggr\}\nonumber\\
&\!\!\!v\equiv0\;(\textrm{constant strain assumption: Taylor/Voigt})\nonumber\\
&\leq\int_V\frac{1}{2}\left\langle \C(\xi)\overline{E},\overline{E} \right\rangle d\xi= \frac{1}{2}\left\langle\overline{E},\int_V \C(\xi)\,d\xi \,\overline{E}\right\rangle=\frac{1}{2}\left| V\right| \left\langle\overline{E},\frac{1}{\left| V\right| }\int_V \C(\xi)\,d\xi \,\overline{E}\right\rangle=\frac{1}{2}\left| V\right| \left\langle\overline{E},\C_{\textrm{Voigt}} \,\overline{E}\right\rangle
\end{align}
it is clear that $\left\langle\C_{\textrm{KUBC}}^{V}\,\overline{E}, \,\overline{E}\right\rangle\leq\left\langle\C_{\textrm{Voigt}} \,\overline{E},\overline{E}\right\rangle$ for all applied loadings $\overline{E}\in\Sym$. On the other hand, it is natural to require as well $\left\langle\cm\,\overline{E}, \,\overline{E}\right\rangle\leq\left\langle\C_{\textrm{Voigt}} \,\overline{E},\overline{E}\right\rangle$, where equality will be obtained if and only if the material on the micro-scale is homogeneous, i.e., $\C(\xi)=\textrm{const}$.
}

\begin{equation}
\frac{1}{2}\left\langle \mathbb{C}_{\textrm{KUBC}}^{V}\,\overline{E},\overline{E}\right\rangle\left|V\left(x\right)\right| =\inf\biggl\{\int_{\xi\in V\left(x\right)}\frac{1}{2}\left\langle \mathbb{C}\left(\xi\right)\left(\sym\nabla_{\xi}v\left(\xi\right)+\overline{E}\right),\sym\nabla_{\xi}v\left(\xi\right)+\overline{E}\right\rangle d\xi\:\Bigl|\:v\in C_{0}^{\infty}\left(V\!\left(x\right),\R^{3}\right)\biggr\}.\label{eq:43}
\end{equation}
The details for this construction will be explained in the next subsection. Combining this with inequality (\ref{uguaglianza energie}) we must have for
all applied loadings $\overline{E}\in\Sym$:
\begin{equation}\label{aa}
\left\langle \cm\,\overline{E},\overline{E}\right\rangle \geq\left\langle \mathbb{C}_{\textrm{KUBC}}^{V}\,\overline{E},\overline{E}\right\rangle .
\end{equation}

\begin{figure}[H]
	
	\begin{centering}
		\includegraphics[scale=0.4]{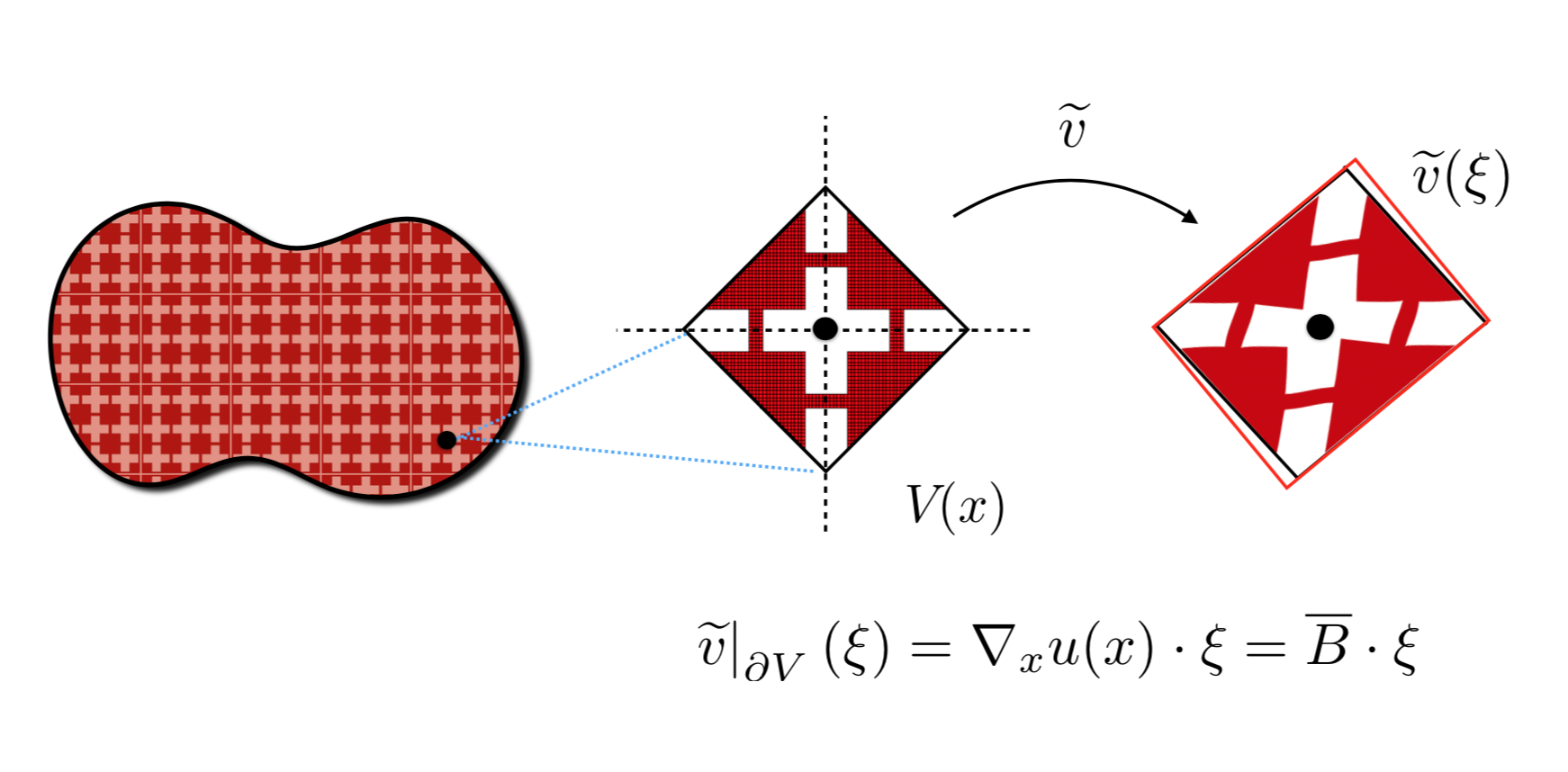}
		\par\end{centering}
	\caption{\label{fig:Affine-Dirichlet-loading}Affine Dirichlet loading (KUBC)
		of the unit-cell $V\!\left(x\right)$. The macroscopic deformation
		state $\nabla_{\xi}u\left(x\right)=\overline{B}$ defines a microscopic
		Dirichlet boundary value problem on the boundary of the microvolume
		$V\!\left(x\right)$ attached to each macroscopic material point $x\in\Omega$.
		Here, we show the superposition of simple shear and elongation.}
\end{figure}

\subsection{The Hill-Mandel energy equivalence for affine Dirichlet conditions}

\label{HM}The Hill-Mandel energy equivalence for KUBC can be obtained easily.
We provide it for the convenience of the reader. On the one hand we have in mechanical equilibrium for linear elasticity

\begin{align}
\int_{\xi\in\partial V}\left\langle \widetilde{v}\left(\xi\right),\sigma\left(\xi\right)\cdot \vec{n}\right\rangle _{\R^{3}}dS & =\int_{\xi\in\partial V}\left\langle \sigma^{T}\left(\xi\right)\cdot \widetilde{v}\left(\xi\right),\vec{n}\right\rangle _{\R^{3}}dS=\int_{\xi\in V}\textrm{div}\left(\sigma^{T}\cdot \widetilde{v}\right)d\xi=\int_{\xi\in V}\textrm{div}\left(\sigma\cdot \widetilde{v}\right)d\xi\nonumber \\
& =\int_{\xi\in V}\left\langle \nabla \widetilde{v},\sigma\right\rangle _{\R^{3\times3}}+\bigl\langle \widetilde{v},\underbrace{\textrm{Div}\,\sigma}_{=0}\bigr\rangle\,d\xi=\int_{\xi\in V}\left\langle \sigma\left(\xi\right),\varepsilon\left(\xi\right)\right\rangle d\xi.\label{eq:(1)}
\end{align}
On the other hand, for KUBC, we have $\left.\widetilde{v}\right|_{\partial V}=\overline{B}\cdot\xi$
and

\begin{align}
\int_{\xi\in\partial V}\left\langle \widetilde{v}\left(\xi\right),\sigma\left(\xi\right)\cdot \vec{n}\right\rangle _{\R^{3}}dS & =\int_{\xi\in\partial V}\left\langle \overline{B}\cdot\xi,\sigma\left(\xi\right)\cdot \vec{n}\right\rangle _{\R^{3}}dS=\int_{\xi\in V}\textrm{div}\left(\sigma^{T}\cdot\left(\overline{B}\cdot\xi\right)\right)d\xi\nonumber \\
& =\int_{\xi\in V}\textrm{div}\left(\sigma\cdot\left(\overline{B}\cdot\xi\right)\right)d\xi=\int_{\xi\in V}\left\langle \nabla\left[\overline{B}\cdot\xi\right],\sigma\right\rangle +\bigl\langle\overline{B}\cdot\xi,\underbrace{\textrm{Div}\,\sigma}_{=0}\bigr\rangle\,d\xi \label{eq:(2)} \\
& =\int_{\xi\in V}\left\langle \overline{B},\sigma\right\rangle d\xi=\big\langle \overline{B},\int_{\xi\in V}\sigma\,d\xi\big\rangle =\left|V\right|\big\langle \overline{B},\frac{1}{\left|V\right|}\int_{\xi\in V}\sigma\,d\xi\big\rangle =\left|V\right|\left\langle \sym\,\overline{B},\overline{\sigma}\right\rangle .\nonumber
\end{align}
Since $\sym\,\overline{B}=\overline{E}=\overline{\varepsilon}$, see
(\ref{eq:37}), taking (\ref{eq:(1)}) and (\ref{eq:(2)}) together
we obtain
\begin{equation}\label{HM2}
\left\langle \overline{\sigma},\overline{\varepsilon}\right\rangle =\frac{1}{\left|V\right|}\int_{\xi\in V}\left\langle \sigma\left(\xi\right),\varepsilon\left(\xi\right)\right\rangle d\xi.
\end{equation}
Summarizing, the Hill-Mandel lemma \eqref{HM2} implies that for KUBC (among other boundary conditions) it holds
that,

\begin{equation}\label{HM3}
\left\langle \overline{\sigma},\overline{\varepsilon}\right\rangle =\frac{1}{\left|V\right|}\int_{\xi\in V}\left\langle \sigma\left(\xi\right),\varepsilon\left(\xi\right)\right\rangle d\xi,\quad\textrm{Div}\,\sigma\!\left(\xi\right)=0,\quad\sigma^{\,T}\!(\xi)=\sigma(\xi),\quad\left.\widetilde{v}\right|_{\partial V}=\overline{B}\cdot\xi,
\end{equation}
where $\overline{\varepsilon},\overline{\sigma}$ are the mean strain
and mean stress, respectively. 

Next, let us assume that on the fine scale we have the linear elastic constitutive
law $\sigma\left(\xi\right)=\mathbb{C}\left(\xi\right)\,\varepsilon\left(\xi\right)$,
where $\mathbb{C}\left(\xi\right)$ is uniformly positive definite.
Then the equilibrium equation $\textrm{Div}\left(\mathbb{C}\left(\xi\right)\,\varepsilon\left(\xi\right)\right)=0$,
$\left.\widetilde{v}\right|_{\partial V}\left(\xi\right)=\overline{B}\cdot\xi$
has a unique (inhomogeneous) solution $\widehat{v}\left(\xi\right)$,
such that $\varepsilon\left(\xi\right)=\sym\,\nabla\widehat{v}\left(\xi\right)$
depends linearly on $\overline{E}=\sym\,\overline{B}$. Thus, the micro-scale Cauchy
stress $\widehat{\sigma}\left(\xi\right)=\mathbb{C}\left(\xi\right)\,\varepsilon\left(\xi\right)$
depends also linearly on $\overline{E}$. On the other hand, it follows
by partial integration that the mean strain tensor satisfies $\overline{\varepsilon}=\frac{1}{\left| V\right| }\int_{\xi\in V}\varepsilon(\xi)\,d\xi=\overline{E}$,
see \eqref{epsbar}, and moreover, that the mean Cauchy stress tensor $\overline{\sigma}=\frac{1}{\left|V\right|}\int_{\xi\in V}\sigma\left(\xi\right)d\xi$
depend also linearly on $\overline{E}$. This implies that there exists
a unique linear mapping  with constant coefficients $\mathbb{C}_{\textrm{KUBC}}^{V}$ such that $\overline{\sigma}=\mathbb{C}_{\textrm{KUBC}}^{V}\,\overline{E}$.
Therefore, using \eqref{HM3}, we must have\footnote{An equivalent, more algorithmic procedure to determine $\mathbb{C}_{\textrm{KUBC}}^{V}$ is obtained as follows. Consider again \eqref{HM3} \begin{equation}
	\left\langle \overline{\sigma},\overline{\varepsilon}\right\rangle =\frac{1}{\left|V\right|}\int_{ V}\left\langle \sigma\left(\xi\right),\varepsilon\left(\xi\right)\right\rangle d\xi=\frac{1}{\left|V\right|}\int_{V}\left\langle \C(\xi)\,\varepsilon\left(\xi\right),\varepsilon\left(\xi\right)\right\rangle d\xi,\quad\textrm{Div}\,\sigma\!\left(\xi\right)=0,\quad\sigma^{\,T}\!(\xi)=\sigma(\xi),
	\end{equation}
and $\widetilde{v}=\overline{\varepsilon}\cdot\xi$ at the boundary. Let us define the corresponding linear solution operator of the linear elastic problem at the micro-scale
$\mathscr{L}(\xi)\cdot\overline{\varepsilon}=\varepsilon(\xi)$,
("localization tensor") and insert this back into \eqref{HM3}. This gives 
{\scriptsize 
\begin{align*}
\left\langle \overline{\sigma},\overline{\varepsilon}\right\rangle &=\frac{1}{\left|V\right|}\int_{V}\left\langle \C(\xi)\,\mathscr{L}(\xi)\cdot\overline{\varepsilon},\mathscr{L}(\xi)\cdot\overline{\varepsilon}\right\rangle d\xi=\frac{1}{\left|V\right|}\int_{V}\left\langle\mathscr{L}(\xi)^T \C(\xi)\,\mathscr{L}(\xi)\cdot\hspace{-4mm}\underbrace{\overline{\varepsilon}}_{\textrm{const. in }\xi}\hspace{-3mm},\overline{\varepsilon}\right\rangle d\xi
=\left\langle\overline{\varepsilon},\underbrace{\left( \frac{1}{\left|V\right|}\int_{V}\mathscr{L}(\xi)^T \C(\xi)\,\mathscr{L}(\xi)\,d\xi\right)}_{\rotatebox[origin=c]{180}{$\defi$}\,\mathbb{C}_{\textrm{KUBC}}^{V}} \cdot\,\overline{\varepsilon}\right\rangle=\left\langle \overline{\varepsilon},\mathbb{C}_{\textrm{KUBC}}^{V}\,\overline{\varepsilon}\right\rangle . 
\end{align*}
}
}
\begin{equation}
\frac{1}{2}\left\langle \mathbb{C}_{\textrm{KUBC}}^{V}\,\overline{E},\overline{E}\right\rangle =\inf\left(\frac{1}{\left|V\right|}\int_{\xi\in V}\frac{1}{2}\left\langle \mathbb{C}\left(\xi\right)\left(\sym\nabla_{\xi}v\left(\xi\right)+\overline{E}\right),\sym\nabla_{\xi}v\left(\xi\right)+\overline{E}\right\rangle d\xi\:\Bigl|\:v\in C_{0}^{\infty}\left(V\!\left(x\right),\R^{3}\right)\right).
\end{equation}

The reader is warned that for classical periodic homogenization, the Hill-Mandel lemma is used with periodic boundary conditions PBC. In this case a unique constant linear mapping $\mathbb{C}^V_{\textrm{PBC}}$ can be obtained accordingly. However, $\mathbb{C}^V_{\textrm{PBC}}$ turns out to be independent of the size and shape of the chosen unit-cell $V$ such that 
\begin{equation}
\mathbb{C}^V_{\textrm{PBC}}\,\rotatebox[origin=c]{180}{$\defi$}\,\mathbb{C}_{\textrm{macro}}
\end{equation}
defines the unique macroscopic homogenized stiffness. It is well known that for large unit-cells $V$, the homogenized apparent stiffness tensor $ \mathbb{C}_{\textrm{KUBC}}^{V}$ approximates $\mathbb{C}_{\textrm{macro}}$ \cite{nemat2013micromechanics,neff2015poincare}, see also \cite{wang2009scale}. In general,  $ \mathbb{C}_{\textrm{KUBC}}^{V}$ is stiffer than $\mathbb{C}_{\textrm{macro}}$ measured in the energy norm \cite{kanit2003determination}. More precisely,
\begin{equation}
\forall\overline{E}\in\textrm{Sym}(3):\quad\big\langle \C^{V_\delta}_{\textrm{KUBC}}\, \overline{E},\overline{E} \big\rangle \geqslant \big\langle \C^{V_{2\delta}}_{\textrm{KUBC}}\,\overline{E},\overline{E} \big\rangle\geqslant\overset{\overset{\delta\fr+\infty}{\longrightarrow}}{\ldots}\geqslant\big\langle \C_{\textrm{macro}}\,\overline{E},\overline{E} \big\rangle,
\end{equation}
where $\delta>0$ is a typical size of the cell $V_\delta$. These hierarchies were first established by \cite{huet1990application}, see also \cite{huet1997integrated,huet1999coupled,sab1992homogenization}.

\begin{figure}[H]
	\begin{centering}
		\includegraphics[scale=0.45]{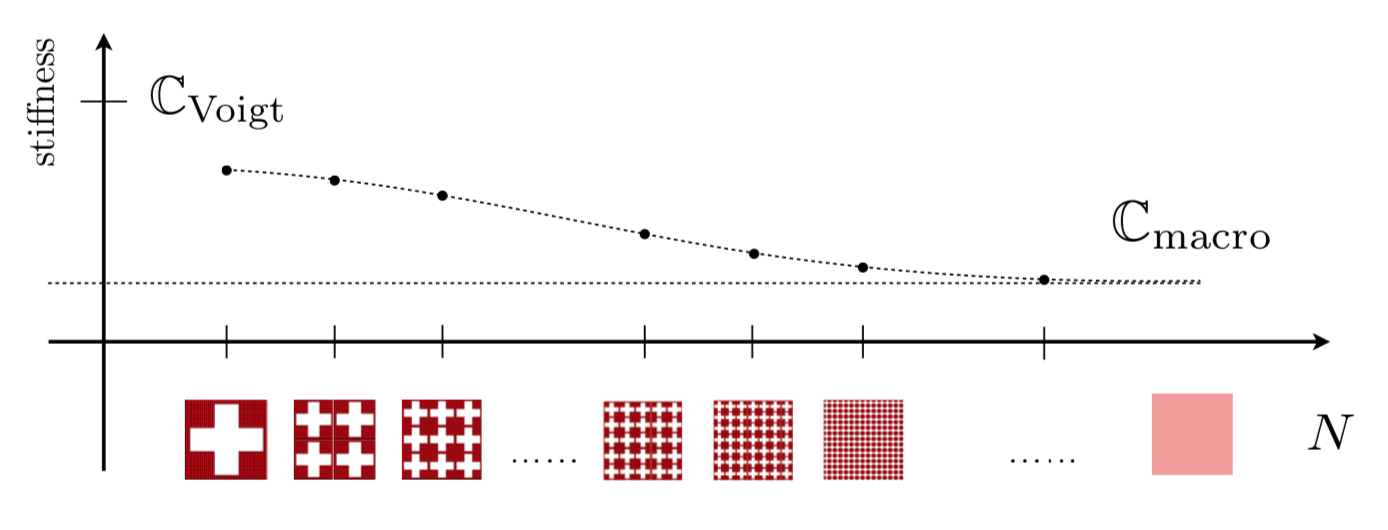}\caption{Qualitative result for the apparent stiffness $ \mathbb{C}_{\textrm{KUBC}}^{V}$, depending on the number of arrays.}
		\par\end{centering}
\end{figure}
 
 Summarizing, both stiffness tensors $ \mathbb{C}_{\textrm{KUBC}}^{V}$ and $\mathbb{C}_{\textrm{macro}}$ involve a homogenization step but they are clearly distinguished and reflect the two-scale nature of the relaxed micromorphic model.
 
 \begin{figure}[H]
 	\begin{centering}
 		\includegraphics[scale=0.40]{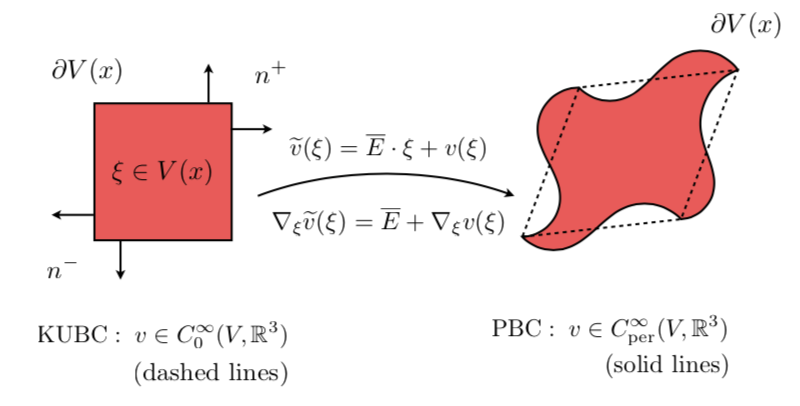}
 		\caption{Difference between affine Dirichlet boundary conditions (KUBC) and periodic Dirichlet boundary conditions (PBC). The fluctuation $v$ is either zero at the boundary (KUBC) or periodic (PBC). The apparent stiffness $ \mathbb{C}_{\textrm{KUBC}}^{V}$ determines $\mathbb{C}_{\textrm{micro}}$ via the Löwner matrix supremum and $\mathbb{C}^V_{\textrm{PBC}}$ determines directly the effective stiffness $\mathbb{C}_{\textrm{macro}}$.}
 		\par\end{centering}
 \end{figure}

\subsection{Neumann's principle and the Löwner bound}

In our given periodic arrangement there are many different possibilities
to choose unit-cells $V$, see Fig.\,\ref{fig:variants-div} in the next section. In the following
we use an \textbf{extended Neumann's principle} \cite{voigt1910lehrbuch,neumann1885vorlesungen} and \cite[p.155]{love1944h},
suitably adapted to our setting:

	\begin{center}\fbox{
		\begin{minipage}[t]{12cm}%
			\medskip{}
			
			\begin{center}
			\textbf{extended Neumann's principle}\tabularnewline[1mm]
			The invariance group of every stiffness tensor of the relaxed \tabularnewline
			micromorphic model $(\ce,\cm,\cc,\mathbb{L})$ must contain the maximal invariance \tabularnewline
			group of the periodic metamaterial\footnotemark.\tabularnewline[1mm]
			\par\end{center}
		%\vspace{1mm}
		\end{minipage}
	}
\par\end{center}
\footnotetext{Here, $\mu\,L^2_c\left\langle\mathbb{L}\,\curl\,P,\curl\,P \right\rangle $ would represent the most general quadratic anisotropic curvature energy in the relaxed micromorphic model, where $\mathbb{L}$ is a fourth-order tensor mapping non-symmetric second-order tensors to non-symmetric second-order tensors.}
Here, the maximal invariance group generated by the periodic metamaterial is the tetragonal group. Therefore the effective stiffness tensor $\cM$ already has tetragonal symmetry. Neumann's principle now requires that $\cm$ must have tetragonal symmetry, too\footnote{$\cm$ could be isotropic nevertheless, since isotropy is a subclass of the tetragonal symmetry.}. Formula \eqref{eq:homog formula} then shows that $\ce$ will also be tetragonal. Moreover, we only consider those apparent stiffness tensors $\mathbb{C}_{\textrm{KUBC}}^{V}$ which are themselves tetragonal. This reduces the number of candidate unit-cells $V$ in \eqref{eq:43} considerably. Finally, we determine a tensor $\cm^0$  in \eqref{aa} by requiring
\begin{equation}
\forall\,\overline{E}\in\Sym:\quad\left\langle \cm^0\,\overline{E},\overline{E}\right\rangle \geq\left\langle \mathbb{C}_{\textrm{KUBC}}^{V}\,\overline{E},\overline{E}\right\rangle \label{eq:CKCm-1}
\end{equation}
for all remaining candidate unit-cell variants $V$ and in addition, any other possible tensor $\widetilde{\mathbb{C}}_{\textrm{micro}}$,
verifying estimate (\ref{eq:CKCm-1}) should satisfy 
\[
\forall\,\overline{E}\in\Sym:\quad\left\langle \widetilde{\mathbb{C}}_{\textrm{micro}}\,\overline{E},\overline{E}\right\rangle \geq\left\langle \cm^0\,\overline{E},\overline{E}\right\rangle .
\]
In this sense $\cm^0$ is optimal. It turns out that $\cm^0$ is a matrix
supremum in the so called Löwner-half-order \cite{burgeth2006mathematical}.

	\begin{center}\fbox{
	\begin{minipage}[t]{16cm}%
		\medskip{}
		
		\begin{center}
			\textbf{Löwner matrix supremum problem}
			\par\end{center}

		Given a family of positive definite symmetric stiffness 
		tensors $\mathbb{C}_{k}\ensuremath{,}k=1,\ldots,n$,  find a positive definite \\ stiffness tensor $\overline{\mathbb{C}}$ such that,
	
		\begin{enumerate}
			\item $\forall\overline{E}\in\textrm{Sym}(3):\quad\left\langle \overline{\mathbb{C}}\,\overline{E},\overline{E}\right\rangle \geq\left\langle \mathbb{C}_{k}\,\overline{E},\overline{E}\right\rangle ,\;k=1,\ldots,n,\quad$
			\textquotedbl $\overline{\mathbb{C}}$ is upper bound in the Löwner order\textquotedbl ,
			\item If $\bigl\langle\widetilde{\mathbb{C}}\,\overline{E},\overline{E}\bigr\rangle\geq\left\langle \mathbb{C}_{k}\,\overline{E},\overline{E}\right\rangle ,\;k=1,\ldots,n$
			then $\bigl\langle\widetilde{\mathbb{C}}\,\overline{E},\overline{E}\bigr\rangle\geq\left\langle \overline{\mathbb{C}}\,\overline{E},\overline{E}\right\rangle\quad$
			\textquotedbl $\overline{\mathbb{C}}$ is least upper bound\textquotedbl .
		\end{enumerate}
\end{minipage}
}
\par\end{center}

\medskip

This is the obtained requirement on $\cm$\footnote{Considering the Voigt upper bound $\C_{\textrm{Voigt}}\defi\frac{1}{\left| V\right| }\int_V\C(\xi)\,d\xi$ as representing the maximal microscopic stiffness is not useful for two reasons:
First, $\mathbb{C}_{\,\textrm{Voigt}}$ will be isotropic and lose the information of the geometry of the microstructure. Second, the actual deformation in any unit-cell will never exhibit constant strain.}.

\bigskip

Gathering our findings, we have obtained the following characterization
\begin{theorem}
	Let $\cM$ be the classical effective elasticity tensor obtained by periodic homogenization for a given periodic microstructure. Assume that the relaxed micromorphic model \eqref{eq:bd33} is chosen as an effective medium to describe the given periodic microstructure. Then, the meso-scale elasticity tensor $\ce$ and the micro-scale elasticity tensor $\cm$ in the relaxed micromorphic model are related by the formula
	\begin{equation}\label{CCCC}
	\mathbb{C}_{\textrm{macro}} =\mathbb{C}_{\textrm{micro}}\left(\mathbb{C}_{\textrm{micro}}+\mathbb{C}_{e}\right)^{-1}\mathbb{C}_{e}. 
	\end{equation}
	Moreover, the microscopic stiffness tensor $\cm$ satisfies the bound
	\begin{equation}\label{Neff3}
	\forall\,\overline{E}\in\Sym:\quad\left\langle \cm\, \overline{E},\overline{E}\right\rangle\geq\left\langle \cm^0\,\overline{E},\overline{E}\right\rangle,  
	\end{equation}
	where $\cm^0$ is the Löwner matrix supremum of the family of apparent stiffness tensors $\mathbb{C}_{\textrm{KUBC}}^{V}$ under affine Dirichlet boundary conditions which are obtained from admissible unit-cells $V$ satisfying the Neumann's principle.
\end{theorem}

\bigskip

\begin{rem}
In the applications \cite{d2019effective}, a first choice is to set $\cm=\cm^0$. However, a fit for the dynamic range may eventually be improved by taking any other positive definite elasticity tensor $\cm$ satisfying estimate \eqref{Neff3}. Identifying $\cm$ as the Löwner supremum is an entirely new approach unique to the relaxed micromorphic model.
\end{rem}

\begin{rem}
In the classical Eringen-Mindlin $W_{\textrm{EM}}-$model \eqref{EM model}, a transparent relation like \eqref{CCCC} is impossible due to the missing proper split of the scale-independent constitutive tensors. On the other hand, the bound \eqref{Neff3} is impossible due to the presence of $\left\|\nabla P \right\|^2 $ instead of $\left\|\textrm{Curl} P \right\|^2$, see inequality \eqref{eq:inf uP}, which fails for $W_{\textrm{EM}}$. This observation highlights the appropriate constitutive assumptions made in the relaxed micromorphic model which allows a rational a priori separation of large and small scale response.
\end{rem}

\subsection{The Löwner matrix supremum in plane strain for tetragonal symmetry}

In the next section, the apparent stiffness tensors $\mathbb{C}_{\textrm{KUBC}}^{V}$ will be determined for the extended Neumann's principle unit-cells $V_i,$ $i=1,\ldots,n$.
In order to distill the missing information on $\cm$ in (\ref{eq:CKCm-1}),
we proceed as follows. Our goal is to find a positive definite tensor
$\cm$ which is the least upper bound of the apparent stiffness of the underlying
microstructure measured in the energy norm. Since the fitting will be done in the 2D-case, we
turn to the planar Voigt representation and the inequality condition
(\ref{eq:CKCm-1}) can be restated as 

\begin{equation}\label{low}
\forall\left(x,y,z\right)^{T}\in\R^{3}:\quad\Bigg\langle \underbrace{\begin{pmatrix}2\widehat{\mu}+\widehat{\lambda} & \widehat{\lambda} & 0\\
\widehat{\lambda} & 2\widehat{\mu}+\widehat{\lambda} & 0\\
0 & 0 & \widehat{\mu}^{*}
\end{pmatrix}}_{\mathbf{\mathbb{C}}_{\mathbf{micro}}}\begin{pmatrix}x\\
y\\
z
\end{pmatrix},\begin{pmatrix}x\\
y\\
z
\end{pmatrix}\Bigg\rangle \geq\Bigg\langle \underbrace{\begin{pmatrix}2\mu+\lambda & \lambda & 0\\
\lambda & 2\mu+\lambda & 0\\
0 & 0 & \mu^{*}
\end{pmatrix}}_{\mathbb{C}_{\mathbf{KUBC}}^{V_i}}\begin{pmatrix}x\\
y\\
z
\end{pmatrix},\begin{pmatrix}x\\
y\\
z
\end{pmatrix}\Bigg\rangle ,\;i=1\ldots n
\end{equation}
where $\left(x,y,z\right)$ represents the planar strain tensor entries $\left(\overline{E}_{11},\overline{E}_{22},\overline{E}_{12}\right)$.
We denote the extended Neumann's principle admissible KUBC-entries as $\mu_{i},\lambda_{i},\mu_{i}^{*},\;i=1,\ldots,n$,
respectively. 

It remains  to obtain numerical values $\widehat{\mu},\widehat{\lambda},\widehat{\mu}^{*}$
such that \eqref{low} is always verified. Since $\mu^{*}$ sits on the diagonal, we must necessarily have that $\mu^{*}\geq\mu_{i}^{*}$ for all $i=1,2,3,4$.
Therefore, the problem is reduced to the $2\times2$ block
\begin{equation}
\forall\left(x,y\right)^{T}\in\R^{2}\;:\;\left\langle \begin{pmatrix}2\widehat{\mu}+\widehat{\lambda} & \widehat{\lambda}\\
\widehat{\lambda} & 2\widehat{\mu}+\widehat{\lambda}
\end{pmatrix}\begin{pmatrix}x\\
y
\end{pmatrix},\begin{pmatrix}x\\
y
\end{pmatrix}\right\rangle \geq\left\langle \begin{pmatrix}2\mu_{i}+\lambda_{i} & \lambda_{i}\\
\lambda_{i} & 2\mu_{i}+\lambda_{i}
\end{pmatrix}\begin{pmatrix}x\\
y
\end{pmatrix},\begin{pmatrix}x\\
y
\end{pmatrix}\right\rangle ,\quad i=1,2,3,4.
\end{equation}
According to the Sylvester-criterion for the difference of positive
definite tensors,
we must have $\widehat{\mu}\geq\mu_{i}$ and $\widehat{\mu}+\widehat{\lambda}\geq\mu_{i}+\lambda_{i}$
for $i=1,2,3,4.$ We choose 
\begin{align}\label{lamlam}
\widehat{\mu} & \defi\max_{i}\left\{ \mu_{i}\right\} , & \widehat{\mu}^{*} & \defi\max_{i}\left\{ \mu_{i}^{*}\right\} ,
&\widehat{\mu}+\widehat{\lambda} & \defi\max_{i}\left\{ \mu_{i}+\lambda_{i}\right\} , & \widehat{\lambda} & \defi\max_{i}\left\{ \mu_{i}+\lambda_{i}\right\} -\widehat{\mu}.
\end{align}
This determines $\mu,\mu^{*},\lambda$ uniquely. In the next section, we have picked the
highlighted values in Table \ref{tab:3} for $\cm$ in (\ref{eq:cccm}).
\section{\label{sec:Static-determination}Elastic parameter identification by numerical homogenization}

For analyses of the considered tetragonal metamaterial in the plane,
the number of independent material constants in linear elasticity
is three for each of the scales. Both
the macro as well as the microscopic elasticity parameters are computed
by numerical homogenization on the unit-cell level, see \cite{michel1999effective,suquet1997effective,suquet1985local}. To that aim we
employ the Finite-Element Heterogeneous Multiscale Method FE-HMM \cite{Eidel,fischer2019convergence},
a two-level finite element method, which is based on asymptotic homogenization
and on the most general Heterogeneous Multiscale Method HMM \cite{weinan2003heterognous}.
A mathematical analysis of FE-HMM for linear elasticity is provided
in \cite{abdulle2006analysis}.

\begin{figure}[H]
	\begin{centering}
		\begin{tabular}{ccc}
			\begin{tabular}{c}
				\tabularnewline
				\includegraphics[scale=0.25]{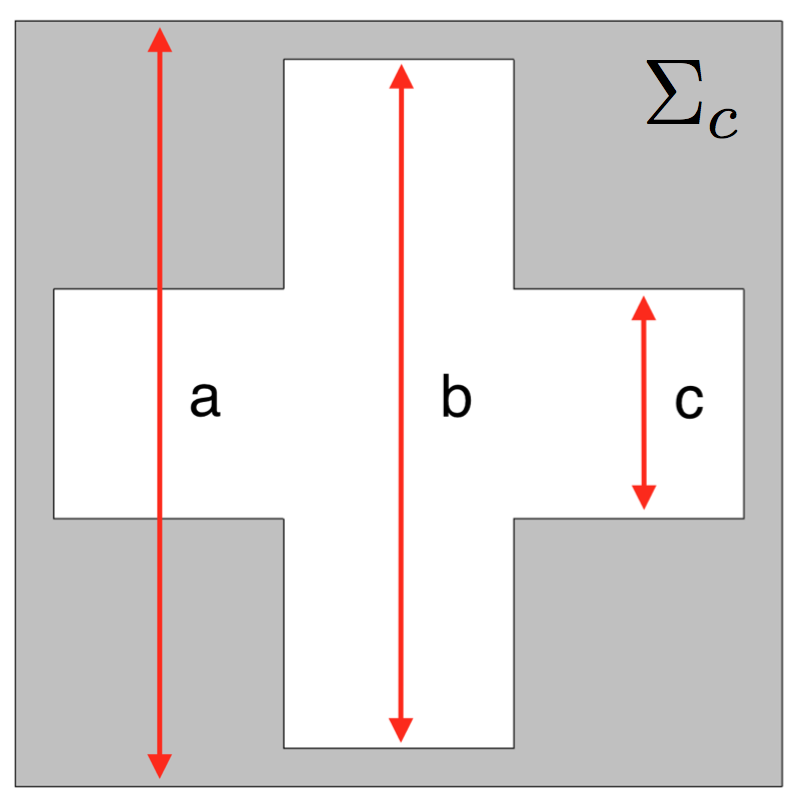}\tabularnewline
				\tabularnewline
			\end{tabular} & %
			\begin{tabular}{c}
				\tabularnewline
				\includegraphics[scale=0.25]{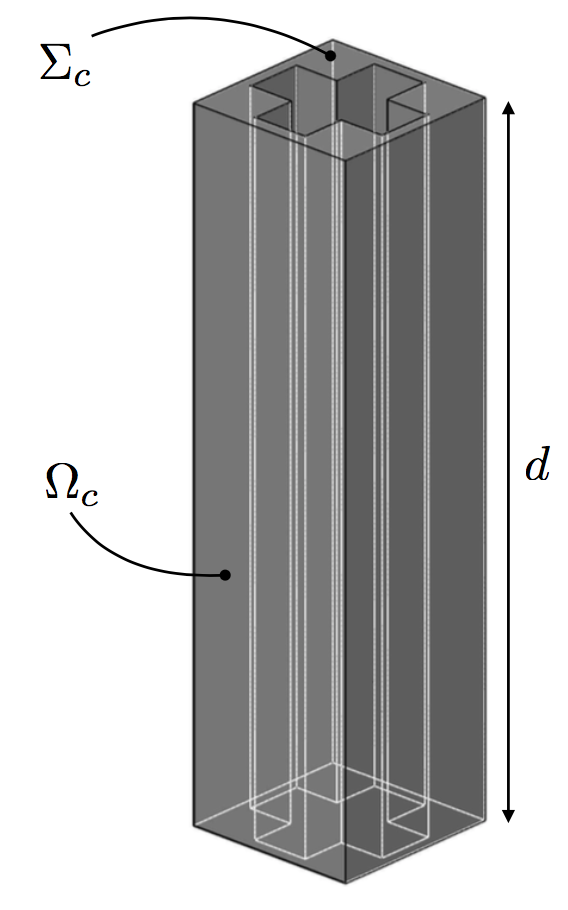}\tabularnewline
				\tabularnewline
			\end{tabular} & %
			\begin{tabular}{c}
				\tabularnewline
				\begin{tabular}{cccc}
					$a$ & $b$ & $c$ & $d$\tabularnewline[1mm]
					\hline 
					\noalign{\vskip1mm}
					$\left[\mathrm{mm}\right]$ & $\left[\mathrm{mm}\right]$ & $\left[\mathrm{mm}\right]$ & $\left[\mathrm{m}\right]$\tabularnewline[1mm]
					\hline 
					\hline 
					\noalign{\vskip1mm}
					$1$ & $0.9$ & $0.3$ & $1$\tabularnewline[1mm]
				\end{tabular}\tabularnewline
				\tabularnewline
				\begin{tabular}{cccc}
					$E$ & $\nu$ & $\mu$ & $\lambda$\tabularnewline[1mm]
					\hline 
					\noalign{\vskip1mm}
					$\left[\mathrm{GPa}\right]$ & $-$ & $\left[\mathrm{GPa}\right]$ & $\left[\mathrm{GPa}\right]$\tabularnewline[1mm]
					\hline 
					\hline 
					\noalign{\vskip1mm}
					$70$ & $0.33$ & $26.32$ & $51.08$\tabularnewline[1mm]
				\end{tabular}\tabularnewline
			\end{tabular}\tabularnewline
		\end{tabular}
		\par\end{centering}
	\caption{Geometry of the unit-cell and elastic parameters for Aluminum.\label{microstr}}
\end{figure}

\subsection{\label{subsec:c_M}Determination of  $\cM-$ classical periodic homogenization}

In mathematical terms, the macroscopic effective stiffness $\mathbb{C}_{\textrm{macro}}$
is obtained by using the classical result of periodic homogenization
(see, e.g., \cite{bensoussan1978asymptotic,boutin2014large,bouyge2002micromechanically}):
\begin{equation}
\dfrac{1}{2}\langle\mathbb{C}_{\textrm{macro}}\,\overline{E},\overline{E}\rangle\left|V\!\left(x\right)\right|:=\inf\left\{ \int_{\xi\in V\!\left(x\right)}\frac{1}{2}\left\langle \mathbb{C}\left(\xi\right)\,\textrm{sym}\left(\nabla_{\xi}v\left(\xi\right)+\overline{E}\right),\textrm{sym}\left(\nabla_{\xi}v\left(\xi\right)+\overline{E}\right)\right\rangle d\xi\:\Bigl|\:v\in C^{\infty}_{\textrm{per}}\!\left(V\!\left(x\right),\R^{3}\right)\right\} ,
\end{equation}
where $\mathbb{C}(\xi)$ is the elasticity tensor of the aluminum
phase or air depending on the position of $\xi$ in the unit-cell\footnote{Here, $x$ is the macro space variable of the continuum, while $\xi$
	is the micro-variable spanning inside the unit-cell.} and $\overline{E}=\sym\nabla u\left(x\right)$ is the applied straining
at the macroscopic point $x$, where the unit-cell $V\!\left(x\right)$
is centered at $x$. For the computation of these macroscopic elasticity
coefficients we use the two-scale finite element method FE-HMM \cite{Eidel}
(see also \cite{michel1999effective,suquet1997effective,suquet1985local})
and we assume that the microproblem is driven under macroscopic plane
strain conditions.

The Lamé constants for the converged solution are obtained for mesh-size
$h=1/2560$~mm (for the geometry of the unit-cell see Figure \ref{microstr}); they are displayed in Table \ref{tab:Homogenized-elasticity-parameter-periodic}.

\begin{table}[H]
\begin{centering}
\begin{tabular}{ccccccc}
\hline 
\noalign{\vskip1mm}
 & geometry  & \multicolumn{2}{l}{boundary conditions} & \multicolumn{3}{l}{elasticity parameters}\tabularnewline[1mm]
\noalign{\vskip1mm}
{\Large{}$\boldsymbol{\mathbb{C}}_{\mathbf{macro}}$} & Fig.\ref{fig:Tetragonal-unit-cells}  &  & plane strain  & $\lambda_{\textrm{macro}}\,${[}GPa{]}  & $\mu_{\textrm{macro}}$\,{[}GPa{]}  & $\mu_{\textrm{macro}}^{\ast}\,${[}GPa{]} \tabularnewline[1mm]
\noalign{\vskip1mm}
\hline 
\hline 
\noalign{\vskip1mm}
 & (a)\textendash (d)  &  & periodic  & $1.738$  & $5.895$  & $0.620$ \tabularnewline[1mm]
\hline 
\end{tabular}
\par\end{centering}
\caption{\label{tab:Homogenized-elasticity-parameter-periodic}Homogenized
macroscopic Lamé constants identified under plane strain and PBC.}
\end{table}

\subsection{\label{subsec:c_m}Determination of $\cm-$ apparent stiffness for affine Dirichlet boundary conditions}

Similar to the macroscopic parameters, the micro set $\cm$ shall be identified
by numerical homogenization. In contrast to the macroscopic scale,
the relaxed micromorphic model imposes conditions at the micro-scale,
which are non-standard, and in particular, in their combination, very
selective as far as the choice of the unit-cells is concerned. The
conditions which are imposed by the relaxed micromorphic model on
choice of the unit-cells and on the boundary conditions read: 
\begin{itemize}
	\item[(i)] they correspond to the case $L_{c}\rightarrow\infty$, a maximal
	zoom into the material, \\[-7mm] 
	\item[(ii)] they represent the maximal stiffness response of the (meta)material
	at the micro-scale, \\[-7mm] 
	\item[(iii)] they reflect the material's overall (tetragonal) symmetry. \\[-7mm] 
\end{itemize}
\medskip{}

The first condition $L_{c}\rightarrow\infty$ is rather vague and
not very selective. The zoom into the single solid phase of the material
however can be ruled out, since the resultant isotropy of aluminum
violates condition (iii).

Condition (ii) alone suggests to consider constant strain conditions,
since they yield the upper bound of stiffness, the Voigt-bound. Under
constant strain assumption however, symmetry information of the microstructure
is lost; numerical homogenization results in an isotropic material
response, $\mu=\mu^{\ast}$, see Table \ref{tab:3}, which again violates
condition (iii). The conclusion is, that conditions (ii) and (iii)
cannot be fulfilled by the constant strain assumption except for the
trivial case of isotropy.

Among the boundary and loading conditions fulfilling the Hill-Mandel
postulate (see section \ref{HM}), affine Dirichlet boundary conditions
are the candidate to estimate the maximal stiffness while preserving
material symmetries. It is well known, that periodic boundary conditions
(PBC) yield less stiff results, and the constant stress assumption
defines the lower bound of stiffness, the Reuss-bound.

\begin{figure}[H]
\begin{centering}
\begin{tabular}{cc}
\includegraphics[scale=0.25]{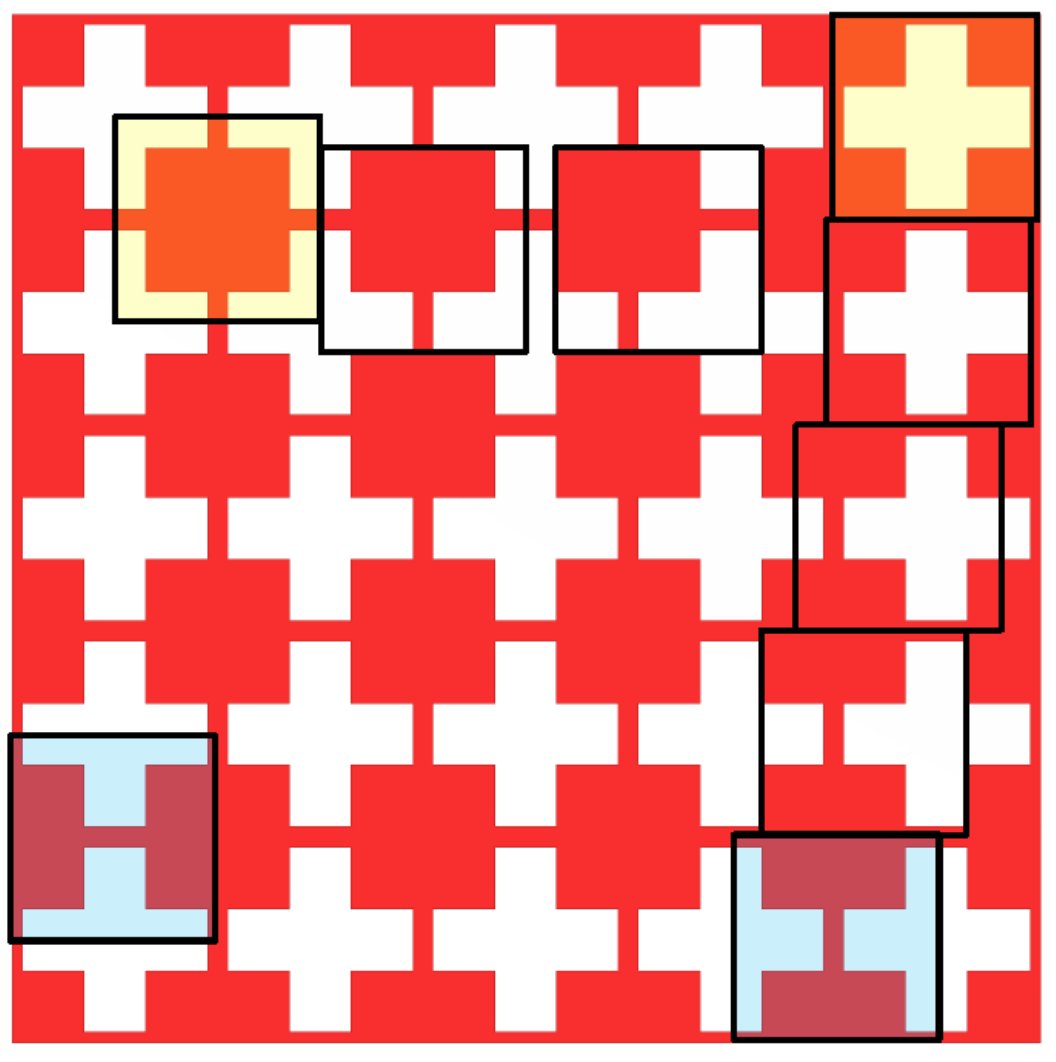} & \includegraphics[scale=0.25]{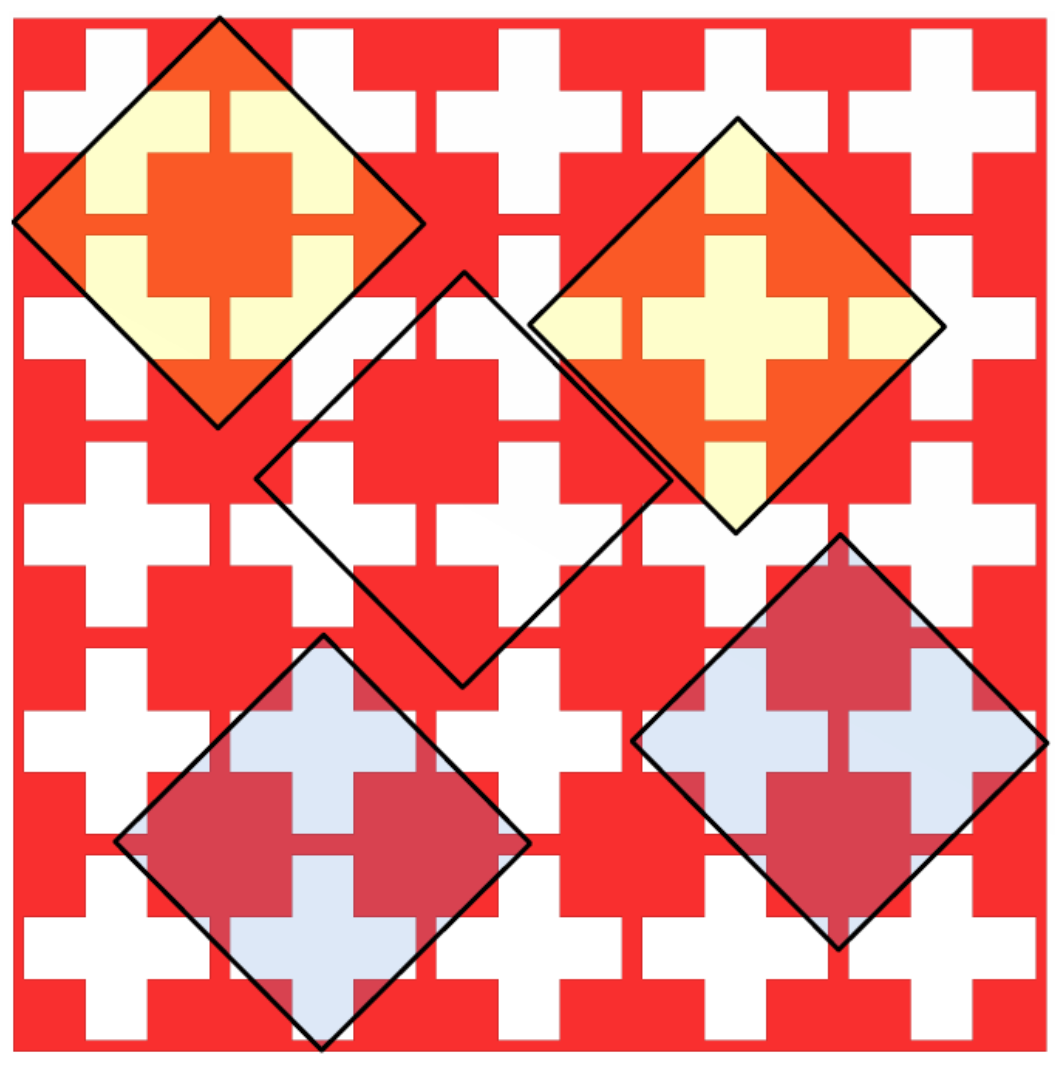}\tabularnewline
(a) & (b)\tabularnewline
\end{tabular}
\par\end{centering}
\caption{\label{fig:variants-div} Identification of candidate unit-cell variants that fulfill the extended Neumann's principle for (a) standard square cells of sidelength $a$ and (b) for rotated square cells of sidelength $a\sqrt{2}$. Yellow/blue shading indicates tetragonal/orthorhombic
symmetries. No shading for cells with only one symmetry axis. Note
that the metamaterial obtained by the infinite repetition of all these
different unit-cells is the same tetragonal metamaterial.}
\end{figure}

There is an infinite number of valid, hence ``equivalent'' unit-cell
variants for the homogenization of periodic media, if PBC are applied\footnote{For a discussion of the non-uniqueness of the unit-cell, see \cite{schroder2014numerical}.}.
Figure~\ref{fig:variants-div} shows some of them for the case of
periodic tessellation based on squares of side length $a$ and based
on rotated squares of sidelength $a\,\sqrt{2}$. Additionally, other
quadrilaterals like rectangles and parallelograms can be used for
valid periodic tessellation as well. They all result in the same material
macroscopic stiffness and they all preserve the tetragonal symmetry,
if PBC are applied to the unit-cell.

The application of affine Dirichlet BCs (KUBC) drastically reduces the above set of periodically
''equivalent'' unit-cells of sidelenghts $a$ and $a\,\sqrt{2}$,
since only four of them capture the tetragonal symmetry under these
boundary conditions. These cells are highlighted in yellow color in
Fig.~\ref{fig:variants-div}, those with two symmetry axes of orthorhombic
materials appear in blue shading. The rest of the displayed unit-cells
exhibit only one symmetry axis. The KUBC render the symmetry criterion
(iii) very selective for unit-cells.

\begin{figure}[H]
\begin{centering}
\begin{tabular}[t]{cccc}
\begin{tabular}{c}
\includegraphics[scale=0.20]{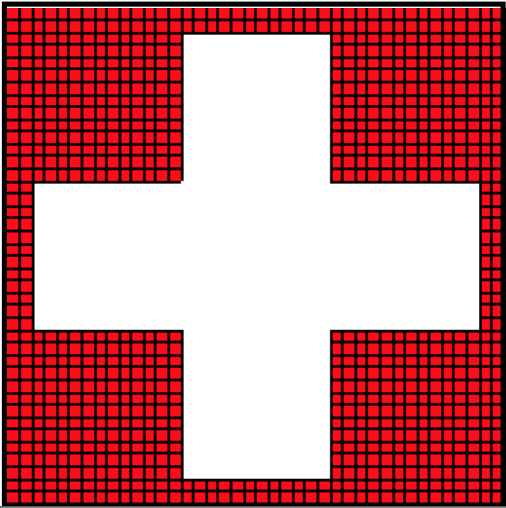}\tabularnewline
\tabularnewline
\end{tabular} & %
\begin{tabular}{c}
\includegraphics[scale=0.2]{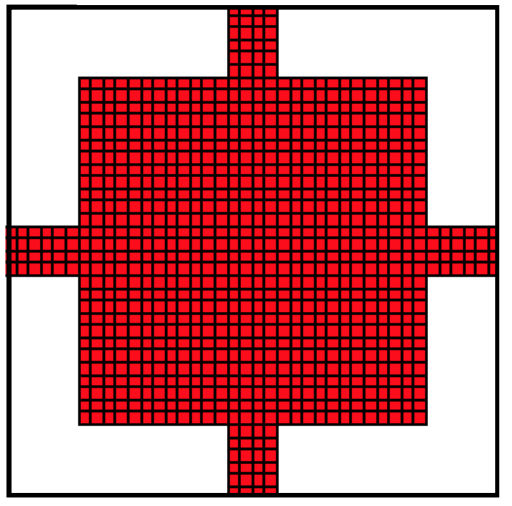}\tabularnewline
\tabularnewline
\end{tabular} & %
\begin{tabular}{c}
\tabularnewline
\includegraphics[scale=0.2]{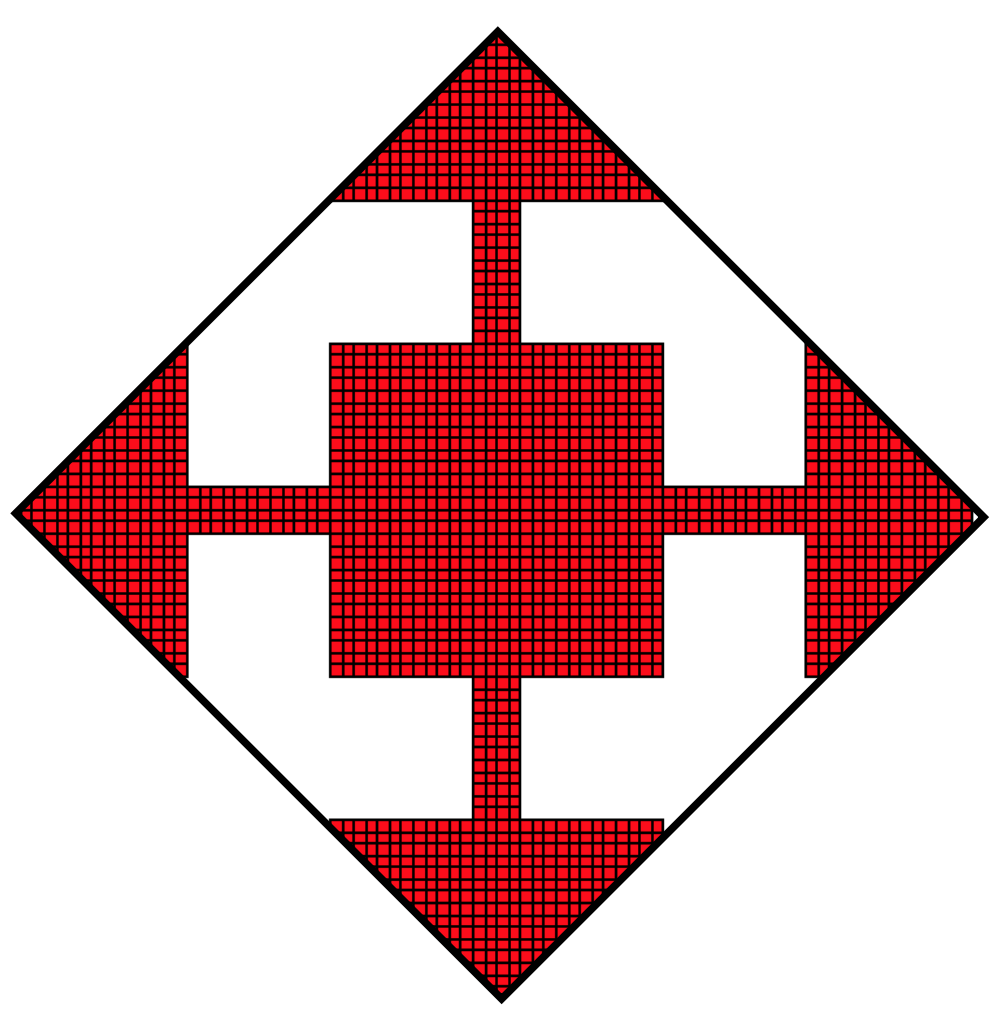}\tabularnewline
\end{tabular} & %
\begin{tabular}{c}
\tabularnewline
\includegraphics[scale=0.2]{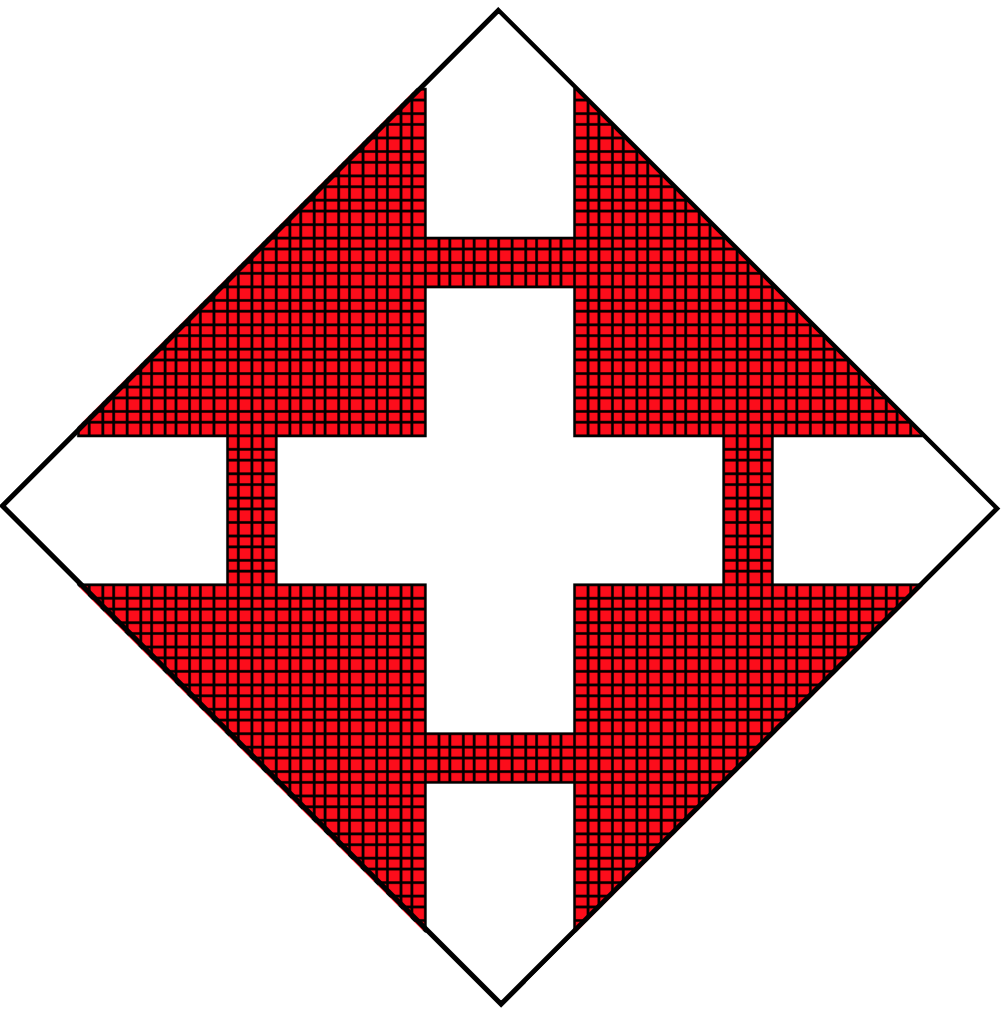}\tabularnewline
\end{tabular}\tabularnewline
(a) & (b) & (c) & (d)\tabularnewline
\end{tabular}
\par\end{centering}
\caption{The four regular square unit-cells respecting tetragonal symmetry.\label{fig:Tetragonal-unit-cells}}
\end{figure}

The requirement for maximal stiffness on top of that condition, however,
does not determine without ambiguity any variant in Fig.~\ref{fig:Tetragonal-unit-cells}
as the single stiffest unit-cell.

\vspace{-2mm}

Nevertheless, from Table \ref{tab:3} we will distill an upper estimate
for the stiffness of the microstructural response. This is done as
follows. For each unit-cell (a), (b), (c), (d) in Figure \ref{fig:Tetragonal-unit-cells},
we have calculated the corresponding apparent stiffness tensor $\mathbb{C}_{\textrm{KUBC}}^{V}$.
We will then determine that positive definite tensor $\cm$ which
has tetragonal symmetry and which satisfies 
\begin{equation}
\forall\,\overline{E}\in\Sym:\quad\left\langle \cm\,\overline{E},\overline{E}\right\rangle \geq\left\langle \mathbb{C}_{\textrm{KUBC}}^{V}\,\overline{E},\overline{E}\right\rangle \label{eq:CKCm}
\end{equation}
for all 4 unit-cells according to Theorem 1.

\begin{table}[H]
\begin{centering}
\begin{tabular}{ccccccc}
\hline 
\noalign{\vskip1mm}
geometry  & \multicolumn{2}{l}{boundary conditions} & \multicolumn{3}{l}{elasticity parameters} & \tabularnewline[1mm]
\noalign{\vskip1mm}
Fig.\ref{fig:Tetragonal-unit-cells}  &  & in $x$-$y$-dir./loading  & $\lambda_{\textrm{hom}}$  & $\mu_{\textrm{hom}}$  & $\mu_{\textrm{hom}}^{\ast}$  & \tabularnewline[1mm]
\hline 
\noalign{\vskip1mm}
(a) \textendash{} (d)  &  & PBC  & $1.738$  & $5.895$  & $0.620$  & \textbf{\Large{}$\boldsymbol{\mathbb{C}}_{\mathbf{macro}}$}\tabularnewline[1mm]
\hline 
\noalign{\vskip1mm}
(a)  &  & KUBC & $4.37$  & $6.242$  & $\boldsymbol{8.332}$  & \tabularnewline[1mm]
\noalign{\vskip1mm}
(b)  &  &  & $2.125$  & $5.899$  & $2.264$  & \textbf{\Large{}$\boldsymbol{\mathbb{C}}_{\mathbf{micro}}$}\tabularnewline[1mm]
\noalign{\vskip1mm}
(c)  &  &  & $\mathbf{5.270}$  & $\boldsymbol{8.927}$  & $4.042$  & \tabularnewline[1mm]
\noalign{\vskip1mm}
(d)  &  &  & $5.981$  & $6.254$  & $4.96$  & \tabularnewline[1mm]
\hline 
\noalign{\vskip1mm}
(a) \textendash{} (d)  &  & constant strain - Voigt  & $28.10$  & $14.47$  & $14.47$  & \textbf{\Large{}$\boldsymbol{\mathbb{C}}_{\mathbf{Voigt}}$} \tabularnewline[1mm]
\noalign{\vskip1mm}
aluminum full  &  & KUBC and PBC  & $51.08$  & $26.32$  & $26.32$  & \tabularnewline[1mm]
\hline 
\end{tabular}
\par\end{centering}
\caption{\label{tab:3}Elastic parameter values in Voigt notation and restricted
to the planar case. Homogenized material parameters {[}GPa{]} for
tetragonal unit-cell variants (a)\textendash (d) in Fig.\ref{fig:Tetragonal-unit-cells}
for KUBC and PBC, for constant strain assumption (Voigt-bound), and
the case of single phase aluminum. Note that the PBC-values are the homogenized effective macroscopic stiffnesses, while the KUBC-values are the homogenized apparent stiffnesses of chosen unit-cells.}
\end{table}

In Table \ref{tab:3}, we also report the computed values of the unit-cell
stiffnesses using constant-strain conditions, as well as the stiffness
of a unit-cell completely filled by aluminum computed both with KUBC
and PBC.

Finally, we remark that in the case of an isotropic unit-cell (full aluminum) the homogenization
results for KUBC and PBC coincide. 

The entries of the Löwner matrix supremum $\cm$ are obtained from
Table \ref{tab:3} as follows\footnote{N.B. the value of $\lambda_{\textrm{micro}}$ is correctly 5.270 and not 5.981 because of the definition of $\widehat\lambda$ in \eqref{lamlam}.}
\begin{equation}
\lambda_{\textrm{micro}}=\mathbf{5.270},\qquad\mu_{\textrm{micro}}=\boldsymbol{8.927},\qquad\mu_{\textrm{micro}}^{\ast}=\boldsymbol{8.332}.\label{eq:cccm}
\end{equation}

In conclusion, the homogenization requirements (i) \textendash{} (iii)
for the set of microparameters along with the Hill-Mandel condition lead
without ambiguity to KUBC and uniquely identify the stiffest microscopic
response in terms of a function of the Lamé-parameters. With $\mathbb{C}_{\textrm{micro}}$
and $\mathbb{C}_{\textrm{macro}}$ in hand, we are able to compute
$\mathbb{C}_{e}$ with formula (\ref{eq:homog formula}).

We have used these values in a diversity of scenarios \cite{ d2019effective,barbagallo2018relaxed,aivaliotis2018low,aivaliotis2019microstructure,aivaliotis2019scattering} to good avail.

\section{Conclusion}

The relaxed micromorphic model is a ``macroscopic continuum''
homogenized model which is able to reproduce the response of the selected
metamaterial including band-gaps with only few material parameters which do not depend
on frequency \cite{d2019effective}. Using the $\curl-$curvature measure $\curl\,P$ instead of the full gradient $\nabla P$ conveys $\cm$ (and a fortiori $\ce$) a scale-independent meaning.  We have mathematically justified that the tensor $\cm$
can be identified with the Löwner-half-order matrix-supremum of suitable apparent
stiffness tensors on the micro-scale. To our understanding this identification
is entirely new. In contrast, $\cM$ follows from standard periodic
homogenization and determines the meso-scale elasticity tensor $\ce$
via the exact micro-macro homogenization formula $
\mathbb{C}_{e}=\mathbb{C}_{\textrm{micro}}\left(\mathbb{C}_{\textrm{micro}}-\mathbb{C}_{\textrm{macro}}\right)^{-1}\mathbb{C}_{\textrm{macro}}$. Summarizing, the salient features of our novel approach of parameter identification
are:
\begin{itemize}
	\item $\cm$ represents the stiffest possible estimate of the linear elastic response of
	any admissible unit-cell under affine Dirichlet boundary
	conditions.
	\item Both $\cm$ and $\cM$ can be determined independently of the characteristic
	length-scale $L_{c}$ of the relaxed micromorphic model.
	\item Both $\cm$ and $\cM$ are readily available by first order numerical homogenization
	 on the unit-cell level.
	\item If the unit-cell is homogeneous,
	then $\cm=\cM$ implies $\ce=+\infty$ and the relaxed micromorphic model
	automatically turns into classical linear elasticity with stiffness
	$\cM$, while the classical Eringen-Mindlin model would turn into a
	second gradient formulation with unbounded stiffness.
	\item For large rigid inclusions in the unit-cell, we have in the limit
	of infinite rigidity that $\cm\fr+\infty$ ($\cM=\ce$), reducing the relaxed micromorphic
	model effectively to a Cosserat model (a model with ``rigid microstructure''), which is sensible.
\end{itemize}

\section{Open problems}	

	We have mathematically justified that the tensor $\cm$
	can be identified with the Löwner-half-order matrix-supremum of suitable apparent
	stiffness tensors on the micro-scale.  On the other hand, $\cM$ follows from standard periodic
	homogenization and determines the mesoscale elasticity tensor $\ce$
	via the micro-macro homogenization formula (\ref{eq:homog formula})
	\[
	\mathbb{C}_{e}=\mathbb{C}_{\textrm{micro}}\left(\mathbb{C}_{\textrm{micro}}-\mathbb{C}_{\textrm{macro}}\right)^{-1}\mathbb{C}_{\textrm{macro}}.
	\]
	For this micro-macro homogenization formula to make sense we need to have that $\cm-\cM$
	is positive definite (and therefore invertible). For this positive definiteness, consider the difference
	\begin{gather}
	\left|V\left(x\right)\right|\frac{1}{2}\left\langle \left(\cm-\cM\right)\overline{E},\overline{E}\right\rangle \geq\left|V\left(x\right)\right|\frac{1}{2}\left\langle \left(\mathbb{C}_{\textrm{KUBC}}^{V}-\cM\right)\overline{E},\overline{E}\right\rangle \qquad\qquad\qquad\qquad\qquad\qquad\qquad\qquad\qquad\qquad\qquad\qquad\qquad\qquad\qquad\qquad\qquad\qquad\qquad\qquad\qquad\qquad\qquad\qquad\qquad\\
	\!\!\!\!\!=\inf\left\{ \int_{\xi\in V\left(x\right)}\frac{1}{2}\left\langle \mathbb{C}\left(\xi\right)\left(\sym\nabla_{\xi}v\left(\xi\right)+\overline{E}\right),\sym\nabla_{\xi}v\left(\xi\right)+\overline{E}\right\rangle d\xi\:\Bigl|\:v\in C_{0}^{\infty}\left(V\!\left(x\right),\R^{3}\right)\right\} \qquad\qquad\qquad\qquad\qquad\nonumber \\
	-\inf\left\{ \int_{\xi\in V\left(x\right)}\frac{1}{2}\left\langle \mathbb{C}\left(\xi\right)\left(\sym\nabla_{\xi}v\left(\xi\right)+\overline{E}\right),\sym\nabla_{\xi}v\left(\xi\right)+\overline{E}\right\rangle d\xi\:\Bigl|\:v\in C^{\infty}_\textrm{per}\left(V\!\left(x\right),\R^{3}\right)\right\} \nonumber 
	=:Q\!\left(\overline{E},\overline{E}\right).\nonumber 
	\end{gather}
	By compactness it would be sufficient for strict positive definiteness
	of $\cm-\cM$ that $Q\!\left(\overline{E},\overline{E}\right)>0\;\forall\,\overline{E}\in\Sym.$
	Although it is easy to see that $Q\!\left(\overline{E},\overline{E}\right)\geq0$
	in general, it remains to investigate under which assumptions on the
	geometry and material of the unit-cell the strict positivity of $Q\!\left(\overline{E},\overline{E}\right)$  can be established
	for other metamaterials. We believe that this is true for microstructures with sufficient contrast in material properties in all directions. This will be subject of further research.
	
	We did not yet approach the determination of the static curvature
	parameters, i.e., the characteristic length scale $L_{c}$ (or the curvature parameters induced by the more general anisotropic quadratic expression $\mu L_c^2\left\langle \mathbb{L}\,\curl\,P,\curl\,P \right\rangle $), where $\mathbb{L}$ is a positive definite fourth-order tensor mapping non-symmetric second-order tensors to non-symmetric second-order tensors). Knowledge of $L_c$ (or $\mathbb{L}$) determines the possible long-range interaction of the microstructure. This task, however, will be  greatly facilitated in future works since we already
	know the scale-independent short-range material parameters $\cm$ and $\cM$ and $\ce$. It
	suffices then, in principle, to perform a range of inhomogeneous boundary
	value problems on different sized clusters of unit-cells (mimicking size-experiments) which activate the curvature terms of the relaxed micromorphic
	model in order to fit $L_{c}$ (or $\mathbb{L}$), now via a suitably generalized Hill-Mandel  energy equivalence condition. This will be the subject of further work.
	
\section*{Acknowledgements}

	Patrizio Neff thanks Samuel Forest (Ecole des Mines, Paris), Geralf Hütter (TU Freiberg) and Jörg Schröder (University of Duisburg-Essen) for helpful discussions. The authors are also indebted to Lev Truskinovsky (ESPCI, Paris) for pertinent remarks which helped improve the paper.

{\footnotesize{}\bibliographystyle{plain}
	\bibliography{library-2}
}{\footnotesize \par}
\end{document}